\documentclass[10pt, aps, prl, twocolumn,
longbibliography, superscriptaddress]{revtex4-2}
\usepackage[colorlinks=true,
linkcolor=blue,
anchorcolor=blue,
citecolor=blue,
urlcolor=blue]{hyperref}
\usepackage{graphicx}% Include figure files
\usepackage{dcolumn}% Align table columns on decimal point
\usepackage{bm}% bold math
\usepackage{mathtools}
\usepackage{orcidlink}
\usepackage{nicematrix}
\usepackage{amssymb, amsmath, amsthm}
\usepackage{graphicx}
\usepackage{listings}
\usepackage{amsfonts}
\usepackage{bbold}
\usepackage{xcolor}
\usepackage{nicefrac}
\usepackage{footnote}
\usepackage{blkarray}
\usepackage[]{appendix}
\usepackage{subfiles}
\usepackage{orcidlink}
\newcommand\ket [1]{\left\vert #1\right>}
\newcommand\bra [1]{\left< #1\right\vert}

\usepackage{xr}  % Add this line to use the xr package
\begin{document}

\author{Hoang-Anh
Le\,\orcidlink{0000-0002-1668-8984}}
\thanks{These authors contributed equally to this work.}
\affiliation{Center for Quantum Nanoscience, Institute for Basic Science, Seoul 03760, Republic of Korea}
\affiliation{Ewha Womans University, Seoul 03760, Republic of Korea}

\author{Saba Taherpour\,\orcidlink{0009-0006-6738-4397}}
\thanks{These authors contributed equally to this work.}
\affiliation{Center for Quantum Nanoscience, Institute for Basic Science, Seoul 03760, Republic of Korea}
\affiliation{Department of Physics, Ewha Womans University, Seoul 03760, Republic of Korea}

\author{Denis Jankovi\'{c}\,\orcidlink{0000-0002-9550-6412}}

\affiliation{Center for Quantum Nanoscience, Institute for Basic Science, Seoul 03760, Republic of Korea}
\affiliation{Ewha Womans University, Seoul 03760, Republic of Korea}

\author{Christoph
Wolf\,\orcidlink{0000-0002-9340-9782}}
\email{wolf.christoph@qns.science}
\affiliation{Center for Quantum Nanoscience, Institute for Basic Science, Seoul 03760, Republic of Korea}
\affiliation{Ewha Womans University, Seoul 03760, Republic of Korea}

\title{ Overcoming limitations on gate fidelity in noisy static exchange-coupled surface qubits}

\begin{abstract}
%An atomic-scale qubit platform has been recently demonstrated using statically exchange-coupled  electron spins on surfaces, where quantum gate operations are implemented through all-electric electron spin resonance. 
Recent experiments demonstrated that the spin state of individual atoms on surfaces can be quantum-coherently controlled through all-electric electron spin resonance. By constructing interacting arrays of atoms this results in an atomic-scale qubit platform. 
However, the static exchange coupling between qubits, limited lifetime and polarization of the initial state, impose significant limits on high-fidelity quantum control. We address this issue using open quantum systems simulation and quantum optimal control theory. We demonstrate the conditions under which high-fidelity operations ($\mathcal{F} \gtrsim  0.9$) are feasible in this qubit platform, and show how the Krotov method of quantum optimal control theory adapts to specific noise sources to outperform the conventional Rabi drivings. Finally, we re-examine the experimental setup used in the initial demonstration of this qubit platform and propose optimized experimental designs to maximize gate fidelity in this platform.  
\end{abstract}

\maketitle

\section{Introduction}

Atomic-scale exchange-coupled spin qubits have recently emerged as a platform for quantum coherent nanoscience and quantum information processing~\cite{Wang2023science,
Wolf2024surface}. This is enabled by scanning tunneling microscopy (STM) which offers bottom-up fabrication of tailored spin systems with atomic precision~\cite{Chen2023harnessing}. When combined with electron spin resonance (ESR) techniques \cite{Baumann2015electron,Seifert2020longitudinal}, STM allows for the coherent manipulation and readout of a single electron spins~\cite{Yang2019coherent, Willke2021coherent}.
%thereby fulfilling most of DiVicenzo's criteria~\cite{Bennett2000quantum, Wolf2024surface}. 
In ESR-STM, an atomically sharp magnetic tip is positioned over a magnetic atom adsorbed on an insulating layer. By applying pulsed radio-frequency (RF) voltages, transitions between spin states can be induced, facilitating coherent control through Rabi oscillations. This setup enables the implementation of arbitrary single-qubit rotations and thus universal control of the spin qubits inside the tunnel junction~\cite{Wang2023universal}. 
To extend coherent control to multiple spins, additional atoms are placed outside of the tunnel junction and exchange interactions between neighboring spins are used to facilitated readout of each spin state by frequency addressing~\cite{Phark2023electric, Phark2023DoubleResonance, Hong2024}. When the applied radio-frequency pulses match the resonant conditions of these transitions, their states can be coherently driven, allowing operation of multi-qubit systems~\cite{Wang2023science, Choi2025electron}. 

%{\color{red} Ref. 12 is replaced with published version.}

Despite the significant advancements in these atomic-scale spin qubits, there are several challenges. A significant one is the limited information about the fidelity of gate operations in previous experiments due to the difficulty of quantum-state tomography in this platform. Another limiting factor is the \textit{always-on} static exchange coupling between spins, which induces a complex time-evolution of the coupled quantum system. Experimentally challenging is also radio-frequency crosstalk, where a RF pulse intended to drive a specific qubit inadvertently affect neighboring qubits when their separation in frequency space is close to the natural linewidth. This phenomenon, also known as cross-resonance, is prevalent in other platforms such as superconducting qubits~\cite{Chad2010}, quantum dot spin
qubits~\cite{Warren2021, Heinz2021} and nuclear magnetic resonance~\cite{Vandersypen2005}. The problem becomes more pronounced when multiple pulses are applied simultaneously. All of this is compounded by rapid decoherence, arising from the interaction of qubits with their environment. This interaction leads to relaxation of spin states, characterized by the energy relaxation time $T_1$, as well as dephasing, characterized by coherence time $T_2$~\cite{Willke2018probing, Paul2017}. In this context, the dynamics of the open quantum system can be described using the Gorini–Kossakowski–Sudarshan–Lindblad (GKSL) master equation which incorporates phenomenological collapse operators to model the dissipative effects of the environment~\cite{Breuer2007open,
Phark2023DoubleResonance}.

Strategies to mitigate some of these challenges have been proposed. For instance, to mitigate RF crosstalk, strategies such as designing qubits with optimized resonant frequency allocation~\cite{Morvan2022frequency}, cross-resonance architecture with lightweight resonator coupler~\cite{Zhao2023crosstalk}, and synchronizing Rabi oscillations have been employed~\cite{Heinz2022}. However, these methods do not universally resolve the issues from decoherence. There also exists strategies to cope with decoherence, such as decoherence-free subspaces~\cite{Lidar1998decoherence,bae2018enhanced}, noiseless subsystems~\cite{viola2001experimental}, dynamical decoupling~\cite{Viola1999dynamical} and  bath-optimal minimal-energy control~\cite{Clausen2010bath}. Yet, these methods are often restricted to specific interactions between system and environment. Quantum optimal control theory (QOCT)~\cite{Brif2010control, Glaser2015training} offers a more general approach to address these challenges. QOCT aims to find the optimal control fields that drives the system's evolution toward a desired unitary transformation~\cite{Palao2002PRL, Palao2003PRA}. This approach has been implemented in various systems such as nuclear magnetic resonance~\cite{Khaneja2005,
Skinner2003application} and molecular spin qudits~\cite{Alberto2022optimal}.

In this work, we are going to employ the Krotov method~\cite{krotov1983iterative, somloi1993, reich2012monotonically} from QOCT for realizing high-fidelity quantum gate operations in static exchange-coupled surface qubits. Our results show that, within certain parameter regimes, high-fidelity control ($\mathcal{F} \gtrsim 0.9$) is feasible even in the presence of energy relaxation and decoherence. In addition, optimized pulses obtained via Krotov method result in significantly improved fidelity by exploiting all degrees of freedom of the control Hamiltonian, and it delivers noise-adapted pulses that under noisy conditions outperform those optimized without noise. Using parameters close to recent experiments, we demonstrate that average fidelities of 0.989 for the NOT gate and 0.887 for the CNOT gate can be achieved in two- and three-spin qubit systems, respectively.
Beside gate operations, we apply the method to entanglement preparation and show that the fidelity approaches the upper bound set by the concurrence and it is largely limited by the thermal population of the initial state. 
Finally, we revisit the experimental setup used in the first demonstration of this static exchange-coupled surface qubits platform and propose improvements guided by our simulations.
Our work demonstrates a promising usage of QOCT in this miniature qubits platform.

\begin{figure*}[t]
    \centering
    \includegraphics[width=\linewidth]{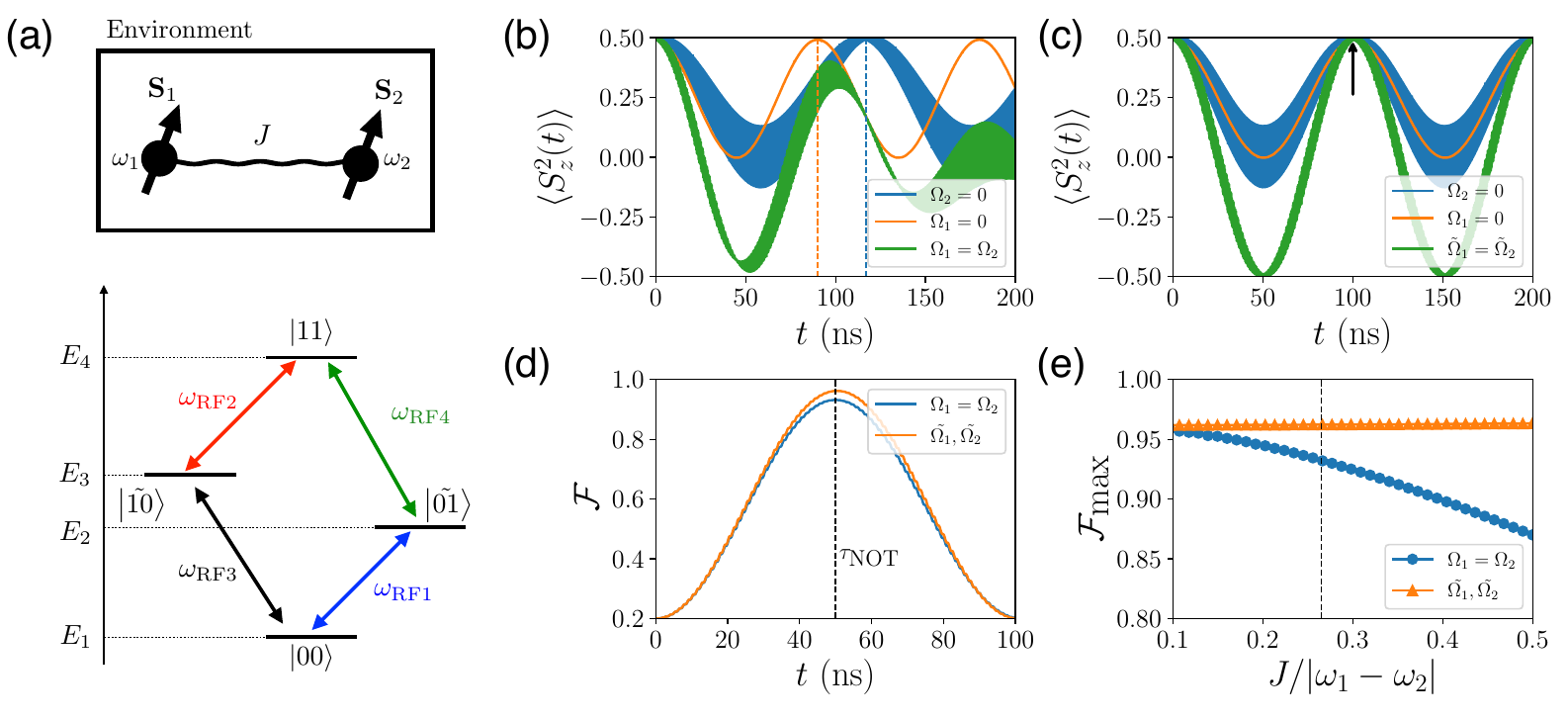}
    \caption{\textbf{Illustration for realization of NOT gate in the closed system.} (a) Schematic of the closed system of two on-surface qubits, i.e. their interaction with surrounding environment is hypothetically ruled out. $J$ denotes their coupling strength and
    $\mathbf{S}_i = \{ S_i^x, S_i^y, S_i^z\}$ is the spin operator vector of $i$th qubit.
    Different Larmor frequencies are realized by different local magnetic fields, $\omega_{1,2} = g \mu_\text{B} B_{1,2}$. 
    The energy diagram and frequencies  corresponding to single-qubit flip are plotted. Analytic forms of eigenenergies $E_1$ to $E_4$ are in Eq.~(\ref{eq:energies}) of the main text. $\omega_{\text{RF}i}$ indicates the transition frequency of the $i$-th transition. 
    (b) Spin evolution of second qubit in lab frame is plotted. The initial state is $1/\sqrt{2}( \ket{00} + \ket{\tilde{10}})$. The blue and orange lines are with respect to transitions
    $\ket{00} \leftrightarrow \ket{\tilde{01}} (\Omega_2=0)$ and $\ket{\tilde{10}} \leftrightarrow
    \ket{11} (\Omega_1=0)$. The parameters of the drift Hamiltonian are $\omega_1 = 20 \pi \text{ GHz}, \omega_2 = 14 \pi \text{ GHz}$, $J = 5 \text{ GHz}$. Driving amplitudes are $\Omega_1 = \Omega_2 = \pi / 50 \text{ GHz}$. Vertical dashed lines mark the peak positions of the corresponding transitions.
    (c) The spin evolution under Rabi rate synchronization shown in
    Eq.~\eqref{eq:phase_sync}. The black arrow indicates the alignment of the two peaks. (d) Fidelity of NOT gate on second qubit as a function of time. 
    $\tau_\text{NOT}$ indicates the
    expected time to realize the NOT gate with given driving amplitudes $\Omega_1$ and $\Omega_2$. (e) Fidelity of the NOT gate with and without using Rabi rate synchronization as a function of coupling strength $J$. Vertical dashed line indicates the coupling strength used in previous figures.}
    \label{fig:fig1_abstract}
\end{figure*}

\section{Results}
\subsection{Static exchange coupled on-surface qubits}

We first consider a generic model for an on-surface qubit system. The drift Hamiltonian describes $N$ spin centers (each with electron spin $S=1/2$) in external magnetic fields and connected via Heisenberg-like exchange coupling. The control Hamiltonian contains the time-dependent magnetic fields at relevant radio-frequencies. The total Hamiltonian reads
\begin{equation}
\begin{aligned}
H(t) = &- \sum_{i=1}^N \omega_i S_i^z + \sum_{i=1}^N \sum_{j>i}^N J_{ij} \mathbf{S}_i \cdot \mathbf{S}_j \\
&+ \sum_i^N \sum_k^{N_k}  \Omega_k \cos \left( \omega_{\text{RF}k} \; \tau_k  + \phi_k \right) \sigma^x_i,
\end{aligned}
\label{eq:general_hamiltonian}
\end{equation}
where the first term describes Zeeman energy with Larmor frequencies $\omega_i$ set by local magnetic fields $\mathbf{B_i}$ {\color{black} (taking the $z$-axis as quantization axis defined by the local magnetic field such that the Larmor term is proportional to $S_i^z$, and $\omega_i = g \mu_B B_i$ with $\mu_B$ Bohr magneton)}, the second term is the Heisenberg isotropic exchange between qubits, which in on-surface qubits is always on and static, i.e. it can not be controllably switched on and off as in other platforms~\cite{He2007switchable, Burkard2023RMP}. %quantum dot and charge + flux qubits 
The final term describes the driving pulses using time-varying perpendicular magnetic fields. $\mathbf{S}_i = \{S_i^x, S_i^y, S_i^z\} = \hbar/2 \{\sigma_i^x, \sigma_i^y, \sigma_i^z\}$ is the spin operator vector of $i$-th qubit, with $S_i^\alpha$ ($\alpha = x, y, z$) as the Pauli spin matrices. $N_k$ denotes the number of simultaneous pulses applied to the system, with $\omega_{\text{RF}k}$, $\Omega_k$, $\tau_k$, being frequency, amplitude, and length of the $k$-th Rabi driving pulse, respectively. We note that we set the phase $\phi_k=0 \; \forall k$ throughout for simplicity but phase differences between driving pulses can in principle affect the gate fidelity (see Supplemental Material Fig.~\ref{fig:angle}). 

\subsection{Limits on gate fidelity in absence of decoherence}
The presence of coupling terms inherently limits the fidelity of quantum gates even in the decoherence-free limit. We use a closed quantum system and discuss the effects of relaxation and decoherence later. To show this, we examine a two-qubit system, as illustrated in Fig.~\ref{fig:fig1_abstract}(a), for a realization of NOT gate on the second qubit. The system's Hamiltonian in Eq.~\eqref{eq:general_hamiltonian} can be simplified as
\begin{equation}
\begin{aligned}
    H_2(t) = &- \omega_1 S^z_1 - \omega_2 S^z_2 + J \mathbf{S}_1 \cdot \mathbf{S}_2 \\ 
    &+ \left[ \Omega_1 \cos (\omega_\text{RF1} t)  + \Omega_2 \cos (\omega_\text{RF2} t) \right] (\sigma_1^x + \sigma_2^x).
\end{aligned}
\label{eq:two_qubit_hamiltonian}
\end{equation}
We first need to solve the eigenstates of the drift Hamiltonian, i.e. the time-independent part of Eq.~\eqref{eq:two_qubit_hamiltonian}. The Heisenberg coupling term hybridizes $\ket{01}$ and $\ket{10}$ states and leaves the other two, $\ket{00}$ and $\ket{11}$, unchanged. Thus, the basis for the drift Hamiltonian is $\left\{
\ket{00}, \ket{\tilde{01}}, \ket{\tilde{10}}, \ket{11} \right\}$, with  
\begin{equation}
\begin{aligned}
\ket{\tilde{01}} &= -\sin \xi_1 \ket{01} + \cos \xi_1 \ket{10},\\
\ket{\tilde{10}} &= \sin \xi_2 \ket{01} + \cos \xi_2 \ket{10},
\end{aligned}
\end{equation}
where
\begin{equation}
\begin{aligned}
\xi_1 &= \arctan \left(\alpha + \sqrt{\alpha^2+1} \right),  \\
\xi_2 &= \arctan \left( -\alpha + \sqrt{\alpha^2+1} \right),
\label{eq:xi1}
\end{aligned}
\end{equation}
and $\alpha \equiv (\omega_1 - \omega_2)/J$ is the dimensionless parameter quantifying the strength of coupling with respect to Larmor frequencies. The eigenenergies in increasing order are 
\begin{equation}
\begin{aligned}
E_1 &= \frac{J}{4} - \frac{\omega_1 + \omega_2}{2},\\
E_2 &= \frac{\omega_1 - \omega_2}{2} \cos{2\xi_1} - \frac{J}{4} (1 + 2 \sin2\xi_1), \\
E_3 &=  \frac{\omega_1 - \omega_2}{2}\cos{2\xi_2}  + \frac{J}{4} (-1 + 2 \sin2\xi_2), \\
E_4 &= \frac{J}{4} + \frac{\omega_1 + \omega_2}{2}.
\end{aligned}
\label{eq:energies}
\end{equation} 
The NOT gate on the second qubit requires transitions $\ket{00} \leftrightarrow \ket{\tilde{01}}$ and $\ket{\tilde{10}} \leftrightarrow \ket{11}$, as illustrated in Fig.~\ref{fig:fig1_abstract}(a). Thus, the corresponding transition frequencies are $\omega_\text{RF1} = E_2 - E_1$ and $\omega_\text{RF2} = E_4 - E_3$, and the driving pulse can be presented as a bichromatic pulse
\begin{equation}
\mathcal{D}(t) \equiv \Omega_1 \cos (\omega_{\text{RF1}}t ) + \Omega_2 \cos (\omega_{\text{RF2}}t ),
\label{eq:driving_pulse}
\end{equation}
where $\Omega_{1(2)}$ are the amplitudes. We note that a controlled NOT gate requires a monochromatic pulse only, that is $\omega_\text{RF2}$ for second qubit as the target one~\cite{Wang2023science}.

The matrix representation of the Hamiltonian in Eq.~(\ref{eq:two_qubit_hamiltonian}) in the eigenstate basis can be written as follows
\begin{widetext}
\begin{equation}
\mathbf{H}_2(t) = 
\begin{bNiceMatrix}[
  first-row,code-for-first-row=\scriptstyle,
  first-col,code-for-first-col=\scriptstyle,
]
& \ket{00} & \ket{\tilde{01}} & \ket{\tilde{10}} & \ket{11} \\
\ket{00} & E_1 & (\cos \xi_1 - \sin \xi_1) \mathcal{D}(t)  & 
(\cos \xi_2 + \sin \xi_2)  \mathcal{D}(t)  & 0 \\
\ket{\tilde{01}} & (\cos \xi_1 - \sin \xi_1) \mathcal{D}(t)  & E_2 & 0 & (\cos \xi_1 - \sin \xi_1) \mathcal{D}(t) 
\\
\ket{\tilde{10}} & (\cos \xi_2 + \sin \xi_2) \mathcal{D}(t)  & 0  & E_3 & (\cos \xi_2 + \sin \xi_2) \mathcal{D}(t)  \\
\ket{11} & 0 & (\cos \xi_1 - \sin \xi_1) \mathcal{D}(t)  & (\cos \xi_2 + \sin \xi_2) \mathcal{D}(t)  & E_4
\end{bNiceMatrix}.
\label{eq:matrix_representation}
\end{equation}
\end{widetext}
There are two potential issues that can limit the fidelity of the NOT gate, as discussed below. The first is the cross-talk between driving pulse frequencies. If the two resonant frequencies are close, $\omega_\text{RF1} \approx \omega_\text{RF2}$, each pulse can off-resonantly drive the other's transition, resulting in a low fidelity NOT gate. 
A solution for this issue is to ensure that the coupling strength $J$ is sufficiently large so that the resonant frequencies are well separated.

\begin{figure*}[t]
    \centering\includegraphics[width=\linewidth]{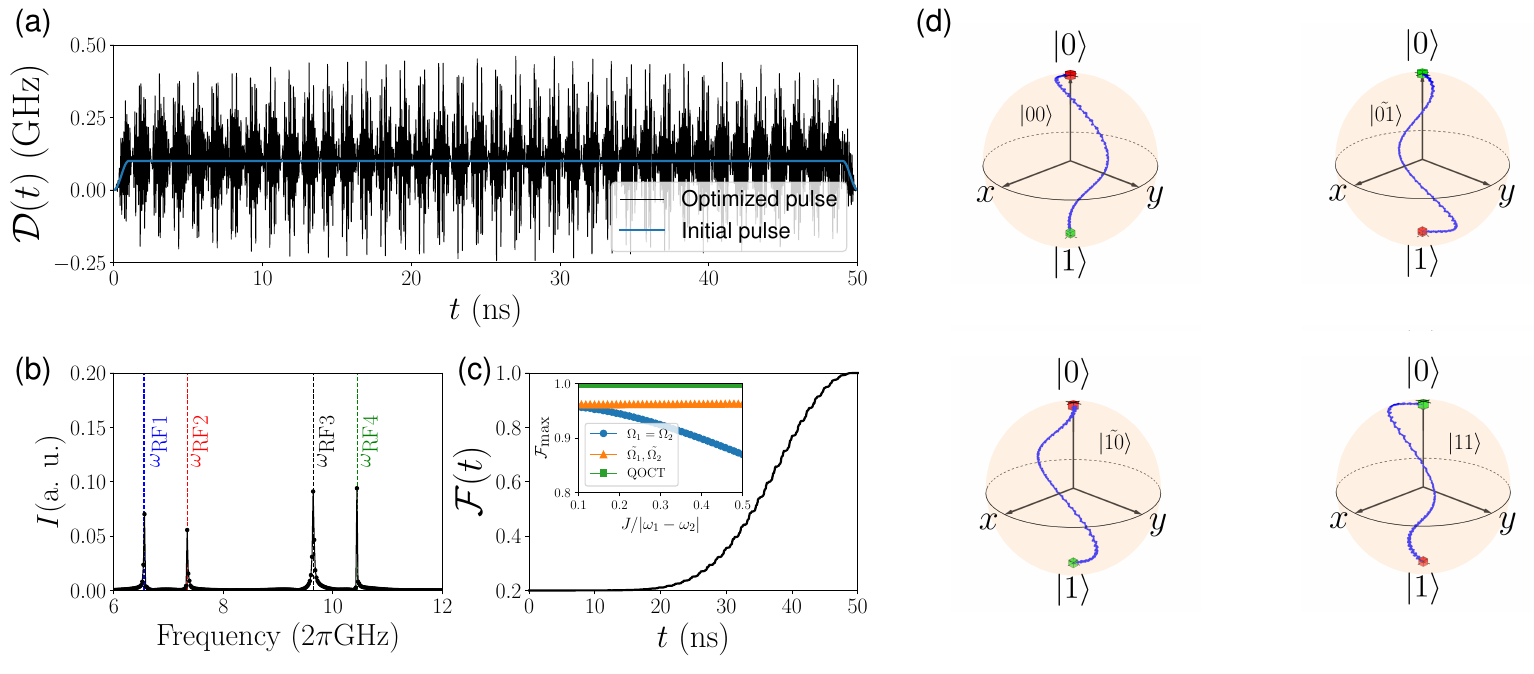}    \caption{\textbf{Realization of a NOT gate on the second qubit using QOCT.} (a) The optimal pulse (black) compared to the initial flattop pulse (blue). The pulse width is set to $\tau = 50$ ns. The drift Hamiltonian parameters are $\omega_1 = 20\pi~\text{GHz}$, $\omega_2 = 14\pi~\text{GHz}$, and $J = 5~\text{GHz}$. The Krotov method parameters are $\lambda_a = 10$ and $N_\text{iter} = 600$. (b) Fourier transform of the optimal pulse. (c) Fidelity of the NOT gate as a function of time, reaching nearly 1 at the end of the optimal pulse. The inset indicates that the unit fidelity can be achieved using Krotov with all values of coupling $J$. (d) Evolution of the second qubit’s state on the Bloch sphere in the interaction picture for all basis states. Red and green cubes represent the initial and final states, respectively.}\label{fig:qoct_no_decoherence}
\end{figure*}

The second challenge is amplitude mismatch between two transitions involved in NOT gate. The NOT gate is realized by setting equal Rabi driving amplitude, $\Omega_1 = \Omega_2$, with a pulse width $\tau_\text{NOT} = \pi / \Omega_1$ for the remote qubit being flipped. However, as implied in Eq.~\eqref{eq:matrix_representation}, the transitions have different effective driving amplitudes induced by the Heisenberg coupling, respectively characterized by
\begin{equation}
\begin{aligned}
\tilde{\Omega}_1 &= \Omega_1 (\cos \xi_1 - \sin \xi_1),\\ \tilde{\Omega}_2 &= \Omega_2(\cos \xi_2 + \sin \xi_2).
\end{aligned}
\end{equation}

The mismatch results in unequal period of transitions depicted in Fig.~\ref{fig:fig1_abstract}(b) and thus worsen fidelity of NOT gates. A solution for this issue is \textit{Rabi rate synchronization}~\cite{Russ2018}, where the Rabi driving amplitudes are adjusted so that the effective amplitude equal to each other $\tilde{\Omega}_1 = \tilde{\Omega}_2$. To achieve Rabi rate synchronization, the two amplitudes should obey the following condition:
\begin{equation}
    \frac{{\Omega}_1}{{\Omega}_2} = \frac{ \cos{\xi_2} + \sin{\xi_2} }{\cos{\xi_1} - \sin{\xi_1}}.
    \label{eq:phase_sync}
\end{equation}
Using the drift Hamiltonian as specified in the caption of Fig.~\ref{fig:fig1_abstract} results in ratio $\Omega_1 / \Omega_2 \approx 1.3$. The two hybridized eigenstates are close with Zeeman basis states, $ \ket{\tilde{01}} \approx 0.992 \ket{01} - 0.129  \ket{10}$ and $ \ket{\tilde{10}} \approx 0.129 \ket{01} + 0.992 \ket{10}$.

The transitions with synchronized Rabi rates are visualized in Fig.~\ref{fig:fig1_abstract}(c) and how it improves the fidelity of NOT gates is shown in Fig.~\ref{fig:fig1_abstract}(d). Note that the above ratio of synchronized amplitude is fully determined via coupling strength with respect to Larmor frequencies.

We extend the analysis with different values of coupling strength $J$. The results are presented in Fig.~\ref{fig:fig1_abstract}(e). As we increase $J$, the mismatch is more severe and leads to lower fidelity. Rabi rate synchronization helps to correct the mismatch and yield nearly constant fidelity over coupling strength. This is because Rabi rate synchronization only addresses the mismatch issue, while there exists other unwanted transitions, $\ket{00} \leftrightarrow \ket{\tilde{10}}$ and $\ket{\tilde{01}} \leftrightarrow \ket{11}$, as seen in Eq.~(\ref{eq:matrix_representation}). Even though the frequencies of these transitions are off-resonant, they are still driven by the pulse, albeit with small coefficients. This results in leakage of population to unwanted states and thus limits the fidelity of NOT gates.

We have seen that even in the simplest case of a two-qubit system, it is not possible to fully
mitigate all the issues raised by exchange coupling, and there exists an upper bound on the
achievable fidelity. The complexity increases exponentially as more qubits become involved in the
dynamics: it becomes difficult to design well separated peaks and to solve the Rabi rate synchronization. 
%A preliminary calculation shows that for 3-qubit system, the maximal fidelity is around 0.52 (see Supplementary Information). 
% In so far, we have limited our analysis to mono- or bichromatic control pulses. To address the
% limitations more thoroughly, we now explore the use of quantum optimal control theory (QOCT) to
% design optimal pulse shapes, which are not necessarily composed of several frequencies, and
% can effectively suppress unwanted transitions.

\begin{figure*}[ht]
    \centering
    \includegraphics[width=\linewidth]{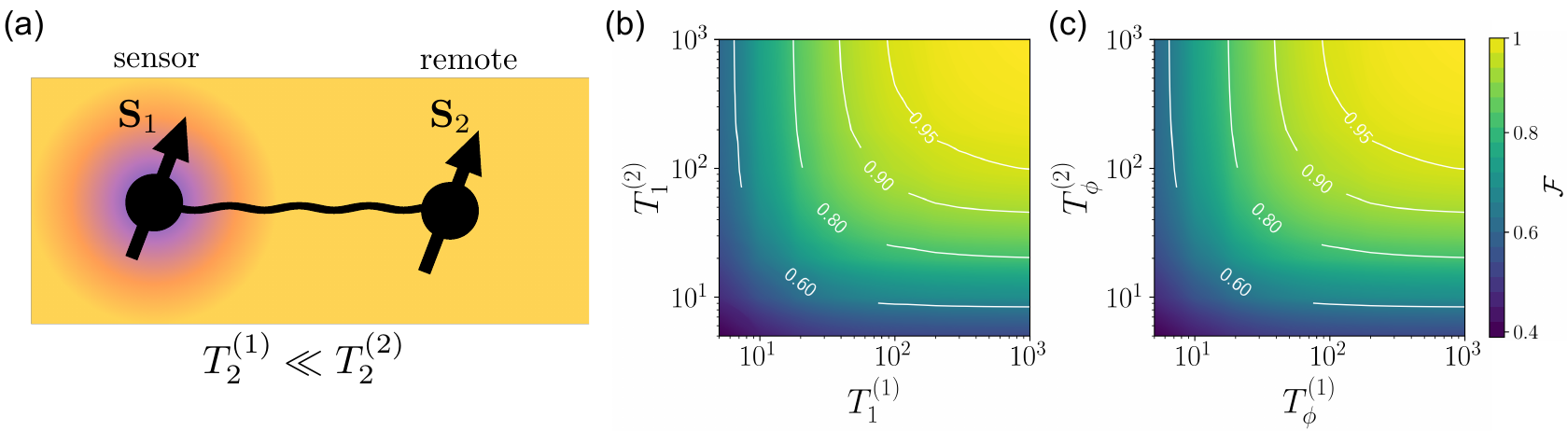}
    \caption{\textbf{Optimized fidelity of a NOT gate on remote qubit with energy relaxation and pure dephasing.} (a) Illustration for the inherent asymmetry in coherence time of
    qubits. The sensor qubit subject to the tunnel current, has a much shorter lifetime whilst the remote qubit is only subject to substrate scattering, i.e. $T_2^{(1)} \ll T_2^{(2)}$. Optimal fidelity of NOT gate on remote qubit obtained with Krotov method for (b) energy relaxation and (c) pure dephasing. Pulse width is $\tau = 10$ ns. The parameters of drift Hamiltonian are $\omega_1 = 20 \pi \text{ GHz}, \omega_2 = 14 \pi \text{ GHz}$, $J = 5 \text{ GHz}$, {\color{black} $T$=0.1 K}.}
    \label{fig:asymmetry_fidelity}
\end{figure*}

\subsection{QOCT in the decoherence-free limit}
\label{sect:qoct_no_decoherence}

{\color{black} Having identified coherent fidelity limits of monochromatic/bichromatic Rabi driving caused by always-on exchange coupling (cross-talk, amplitude renormalization, and leakage), we now ask whether these errors can be removed by optimizing the full control waveform. We first answer this in the decoherence-free (closed-system) limit, where any remaining infidelity is purely coherent, before moving to optimization in the presence of noise. Using the same drift Hamiltonian as above, we apply the Krotov method (see Methods for details) to realize a NOT gate on the second qubit at fixed gate time and with near-unit fidelity.}

%In this section, we employ Krotov method from QOCT (see Methods) to design optimal pulse shapes for realizing a NOT gate on the second qubit. We apply the method on the identical drift Hamiltonian as in the previous section. %{\color{black} It is worth to note that the QOCT framework is fully general and can be applied to arbitrary drift Hamiltonian.}

To set an identical condition for comparison, we set the width for the optimal pulse to be $\tau = 50$ ns, which is the same as the pulse width for Rabi rate synchronization method. We use a flattop pulse as the initial guess for optimization. The Krotov method parameters are $\lambda_a = 10$ and $N_\text{iter} = 600$ (see Methods for detailed explanation). The optimal pulse obtained after 600 iterations is presented in Fig.~\ref{fig:qoct_no_decoherence}(a). It is more relevant to analyze the frequency domain obtained via Fourier transform of the optimal pulse, as shown in Fig.~\ref{fig:qoct_no_decoherence}(b). The spectrum reveals four pronounced peaks located at $\omega_\text{RF1,2,3,4}$, which are resonant frequencies corresponding to single-qubit flips of the drift Hamiltonian (see Fig.~\ref{fig:fig1_abstract}(a)). This is in contrast to the Rabi rate synchronization method, where the spectrum obviously has only two peaks at $\omega_\text{RF1}$ and $\omega_\text{RF2}$. One can consider that the initial flattop pulse comprises all frequencies, and the optimization process with Krotov method selects the relevant ones. The optimal pulse found via Krotov method shows a realization of the NOT gate with unit fidelity (up to numerical error), as shown in Fig.~\ref{fig:qoct_no_decoherence}(c). In addition, the inset indicates that the optimal pulse can achieve nearly perfect fidelity for any coupling strength $J$ {\color{black} as long as the Larmor frequency difference is non-zero and $J$ is large enough to clearly separate the individual transition energies (see also Supplemental Material, Fig.~\ref{fig:J_omega})}. This result suggests that the Krotov method utilizes all degrees of freedom in the control Hamiltonian to achieve high-fidelity quantum gates.

We further illustrate the spin evolution of the second qubit, corresponding to four initial states, on the Bloch sphere in Fig.~\ref{fig:qoct_no_decoherence}(d). The spin trajectory is computed from the reduced density matrix, with fast oscillations removed using interaction picture (see Methods for detail). It is worth to notice that the vector of final state is slightly smaller than unit length due to the partial mixing of eigenstates.

To support our argument on utilization of all degrees of freedom, we have also analyzed two different cases. One case is the realization of a CNOT gate, which, in principle, requires only $\omega_\text{RF2}$. The other case is with Ising interaction where there are no transverse interactions to alter driving amplitude felt by the system. In other words, a situation in which the NOT gate can be realized with unit fidelity even by implementing a bichromatic pulse $\omega_\text{RF1}$ and  $\omega_\text{RF2}$ with equal amplitude. In both cases, the optimal pulse is still able to achieve nearly perfect fidelity and its spectrum still shows four pronounced peaks at the resonant frequencies. The details of these two cases are provided in the Supplemental Material (Fig.~\ref{fig:closed_Ising}). 
{\color{black} In addition, we have also employed a gradient-based method of QOCT, GRadient Ascent Pulse Engineering (GRAPE)~\cite{Khaneja2005}, to optimize the NOT gate in closed systems (Fig.~\ref{fig:closed_GR}). In closed systems GRAPE and Krotov both achieve near unity fidelity. We found that the spectrum of the optimal pulse obtained via GRAPE is strongly dependent on the initial guess pulse and often uses frequencies that don't correspond to the natural resonances of the system, making an analysis in the context of this work difficult (Fig.~\ref{fig:closed_GR_init} and Fig.~\ref{fig:closed_GR_init_fid}). The details are discussed in the Supplemental Material.}

\subsection{QOCT in the presence of decoherence}
\label{qoct_decoherence}

\begin{figure*}[ht]
\centering
\includegraphics[width=\linewidth]{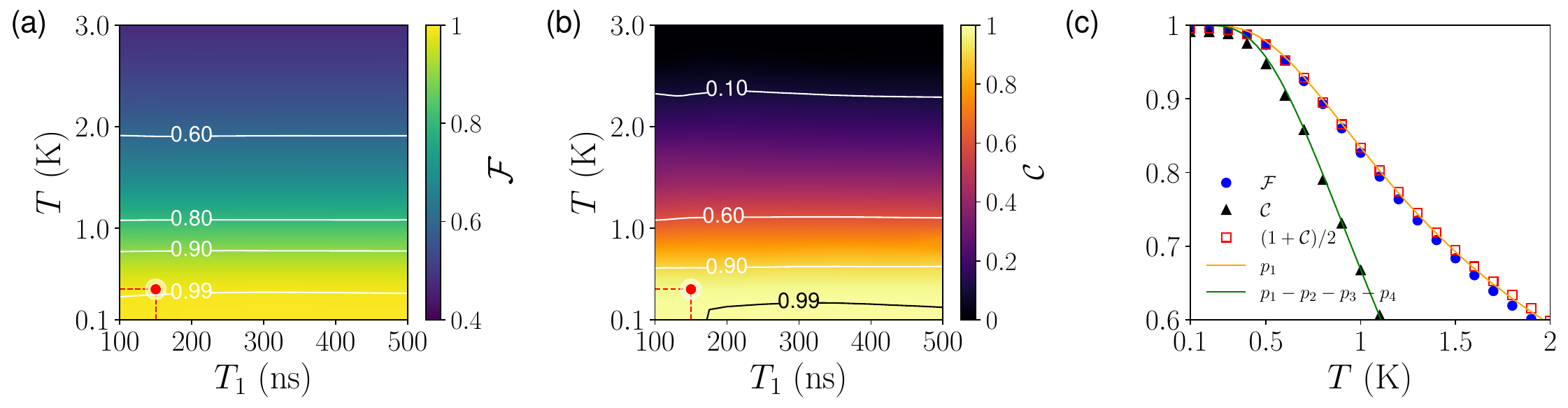}
\caption{\textbf{ {\color{black} Optimized state-to-state preparation for a} maximally entangled two-qubit state.}
(a) Fidelity and (b) Concurrence of the final state as a function of relaxation time $T_1$ and temperature $T$, assuming symmetric lifetime $T_1^{(1)} = T_1^{(2)}$. We employ Krotov method for state-to-state transition from thermal equilibrium state at temperature $T$ to Bell state $\ket{\Phi^+}$.
(c) Comparison of $\mathcal{F}$ and $\mathcal{C}$ with the Boltzmann factors of the eigenstates of the drift Hamiltonian at different temperatures. Filled blue circles and filled black triangles indicate the fidelity and concurrence at $T_1 = 200$ ns, respectively. Empty red squares indicate the upper bound of fidelity for a given concurrence (see main text for detail).
The parameters of the drift Hamiltonian  are $\omega_1 = 20\pi~\text{GHz}$, $\omega_2 = 14\pi~\text{GHz}$, and $J = 5~\text{GHz}$. A flattop pulse is used as the initial guess and the Krotov method parameters are $\lambda_a = 0.01 $ and $N_\text{iter} = 200$. The pulse width is $\tau = 50 \text{ ns}$.}
\label{fig:qoct_entanglement_open}
\end{figure*}

{\color{black}The results for the closed-system establish that, once the full waveform is optimized, the coherent limitations imposed by always-on exchange coupling can be largely removed, enabled by a control spectrum that resembles the relevant system resonances. In experiments, however, gate performance is ultimately limited not by coherent leakage  but by irreversible coupling to the environment during the control pulse. To estimate the limitations imposed by the environment we extend the Krotov optimization to open-system dynamics and quantify how energy relaxation and pure dephasing reshape both the achievable fidelities and the structure of the optimal control fields. We model the dynamics with a GKSL master equation and consider the two dominant channels, energy relaxation and pure dephasing, separately to isolate their individual impact on the achievable fidelity.}

%In this section, we investigate the gate fidelity obtained via Krotov method in the presence of decoherence. In the actual systems, the atomic spin qubits lose coherence due to their coupling with the surrounding environments. Two main decoherence mechanisms are considered: energy relaxation and pure dephasing. The dynamics of the open quantum system is described using a GKSL master equation (see Methods for detail).

The relaxation and pure dephasing processes are characterized by relaxation time $T_1$ and pure dephasing time $T_\phi$, respectively. The overall decoherence rate, which is the inverse of coherence time $T_2$, for a particular $k$th spin can be expressed as
\begin{equation}
    \frac{1}{T_2^{(k)}} = \frac{1}{2 T_1^{(k)}} + \frac{1}{T_\phi^{(k)}}.
\end{equation}
We are going to investigate the two decoherence mechanisms independently. For pure dephasing, we set $T_1^{(k)} \to \infty$, while for relaxation, we set $T_\phi^{(k)} \to \infty$. The drift Hamiltonian parameters are set to be identical as in the previous section. 

In addition, it is worth to note that in the ESR-STM platform, typically, the qubits have \textit{inherently asymmetric energy relaxation rates}, as illustrated in Fig.~\ref{fig:asymmetry_fidelity}(a). In such a system, the sensor spin is subject to the tunneling electrons which is necessary to facilitate the readout but inevitably leads to  $T_2^\text{(1)} \ll T_2^\text{(2)}$. 

% In the context of ESR-STM one can define two paths for energy relaxation. Generally, all spins are subject to energy relaxation to the substrate whilst the spin in the STM junction is additionally subject to relaxation from the current which in general is much faster than relaxation due to substrates scattering. In accordance with literature we refer to the spin in the junction as sensor spin whilst the remaining spins are remote spin. In previous works it has been found that pure dephasing does not seem to play a significant role such that $T_2=2T_1$ for all spins. However, that only applies to remote qubits, while the sensor qubit, which is subject to the junction current, typically has a much shorter coherence time due to the junction current. It establishes an \textit{inherent asymmetry in the coherence time of qubits}, as illustrated in Fig.~\ref{fig:asymmetry_fidelity}(a).

Figures~\ref{fig:asymmetry_fidelity}(b) and (c) show the optimized fidelity of a NOT gate on a remote qubit with respect to relaxation and dephasing mechanisms, respectively. A representative value of pulse width $\tau = 10$ ns is used. {\color{black} The optimal fidelity in both mechanisms share expected features that fidelity increases as the coherence time increases. However, the properties of optimized pulses show subtle differences between these two cases, which will be discussed in the following section.} In the limit of infinitely long coherence time, we recover unit fidelity as in closed systems. As the asymmetry in coherence time is imposed, the optimal fidelity is bounded by the coherence time of sensor qubit, i.e. $T_1^{(1)}$ or $T_\phi^{(1)}$. The upper bound of optimal fidelity depends on the relative coherence time of sensor qubit with respect to pulse width. For instance, if $T_1^{(1)} \gg \tau$ or $T_\phi^{(1)} \gg \tau$, the optimal fidelity can be nearly 100\%. However, if $T_1^{(1)}$ is comparable to $\tau$, the optimal fidelity is significantly reduced. 

%{\color{black}This motivates considering an additional, distinct limitation relevant to state preparation in the platforms considered in this study: the finite thermal polarization of the initial state.}

%The results suggest that optimal pulse should be designed with a duration shorter than the coherence time of sensor qubit, in order to achieve a high-fidelity quantum gate.

%\subsection{State-to-state optimization for entanglement preparation}
\subsection{QOCT at finite temperatures: impact on entanglement preparation}

{\color{black}The discussion so far focused on gate implementation, where decoherence during the pulse limits the achievable average gate fidelity. A complementary task is the preparation of entangled states, where limitations can arise already from the temperature-limited purity of the initial state. To understand the limitations imposed by finite temperature we simulated state-to-state transfer to create maximally entangled Bell states starting from thermally mixed states~\cite{nielsenchuangbook}. We implemented this state-to-state transfer using the same Krotov-based optimization~\cite{krotov1983iterative, somloi1993}, (see Methods). We separately vary $T$ and the relaxation time $T_1$ to disentangle the roles of thermal population and finite lifetime during the pulse.} State-to-state transfer is a natural choice to achieve the optimized pulse, as its cost function is less constrained compared with one of gate optimization~\cite{nosrati2020robust,Yu2021entanglement}.

The initial state is a density matrix at thermal equilibrium of temperature $T$, of which is given by
\begin{equation}
\rho_0 = \frac{e^{-H_0 / k_B T}}{\text{Tr} \left[ e^{-H_0 / k_B T} \right]} = \sum_{i=1}^{4} p_i \ket{\psi_i} \bra{\psi_i},
\end{equation}
where $k_B$ is the Boltzmann constant, $\ket{\psi_i}$ are the eigenstates $\{ \ket{00}, \ket{\tilde{01}}, \ket{\tilde{10}}, \ket{11} \}$ of the drift Hamiltonian $H_0$ in Eq.~\eqref{eq:two_qubit_hamiltonian}, and $p_i = \exp \left(-E_i / k_B T \right) / \sum_{j=1}^{4} \exp \left(-E_j / k_B T\right)$ are the Boltzmann factors with $E_i$ being the corresponding eigenenergies (see Eq.~\eqref{eq:energies}).
% $H_0$ denotes the drift Hamiltonian which critically influences the state polarization at a given temperature. This opens additional venues for optimization of the quantum system which will not be discussed in this work. 
The target state is one of the four Bell states  $\ket{\Phi^+} = (\ket{00} + \ket{11})/\sqrt{2}$.
In addition to the fidelity between Bell state and the final state (obtained via Krotov method), we use concurrence $\mathcal{C}$ of the final state as an additional metric. Concurrence quantifies entanglement of two qubits: concurrence of fully separated states is zero and concurrence of maximally entangled states is one. 

\begin{figure*}[ht]
    \centering
    \includegraphics[width=\linewidth]{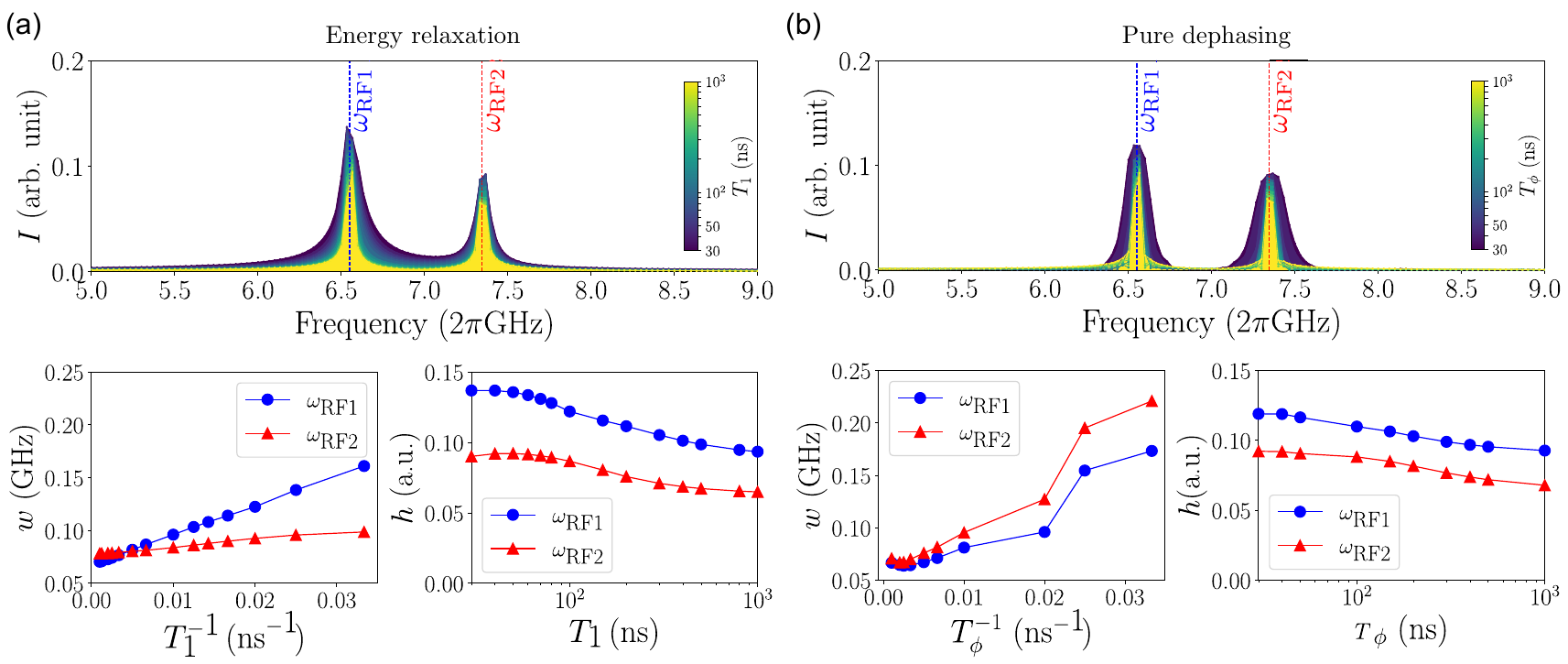}
    \caption{\textbf{Properties of the spectrum of the optimized pulse for a NOT gate on remote qubit with respect to (a) energy relaxation and (b) pure dephasing.} The evolution of the spectrum with respect to corresponding coherence time, and its width and height at $\omega_\text{RF1}$ and $\omega_\text{RF2}$ are shown. The parameters of drift Hamiltonian are
    $\omega_1 = 20 \pi \text{ GHz}, \omega_2 = 14 \pi \text{ GHz}$, $J = 5 \text{ GHz}${\color{black} , $T = 0.1~\text{K}$}. The pulse width is fixed with $\tau = 30$ ns. A flattop pulse is used as the initial guess and $N_\mathrm{iter} =  100$.
    }
    \label{fig:spectrum_vs_relaxation}
\end{figure*}

Figures~\ref{fig:qoct_entanglement_open}(a) and (b) present the results of preparation of Bell state $\ket{\Phi^+} = (\ket{00} + \ket{11})/\sqrt{2}$ as a function of lifetime of each qubits and temperature, using optimization for state-to-state transfer. We observe that both fidelity and concurrence do not depend on lifetime and are mainly determined by temperature. High fidelity and concurrence are achieved at low temperatures. This result holds for other Bell states, as summarized in Tab.~\ref{tab:bell_states}.

The strong dependence of fidelity and concurrence on temperature can be understood by analyzing the Boltzmann factors of the eigenstates of the drift Hamiltonian at different temperatures, as shown in Fig.~\ref{fig:qoct_entanglement_open}(c). 
At low temperatures, the initial state is predominantly in the ground state $\ket{00}$, where Krotov method can efficiently drive the system to the target Bell state $\ket{\Phi^+}$. This claim is supported by the fact that fidelity and the population of ground state $p_1$ exhibit similar trends as temperature varies. In addition, the fidelity obtained from the Krotov method is close to its upper bound for a given concurrence~\cite{Verstraete2002Entfidelity}
\begin{equation}
    \mathcal{F} \leq \frac{1 + \mathcal{C}}{2}.
\end{equation}
The match between concurrence and $p_1 - p_2 - p_3 - p_4$ is a derived result from the above equation, using the normalization condition $\sum_i^4 p_i = 1$.

\begin{table}[h!]
    \renewcommand{\arraystretch}{1.5}
    \centering
    \begin{tabular}{|c|c|c|}
        \hline 
        \textbf{Bell State} & $\mathcal{F}$ & $\mathcal{C}$ \\
        \hline
        $\ket{\Phi^+} = \frac{1}{\sqrt{2}}(\ket{00} + \ket{11})$ & 0.9849 & 0.9712 \\
        $\ket{\Phi^-} = \frac{1}{\sqrt{2}}(\ket{00} - \ket{11})$ & 0.9878 & 0.9760 \\
        $\ket{\Psi^+} = \frac{1}{\sqrt{2}}(\ket{\tilde{01}} + \ket{\tilde{10}})$ & 0.9816 & 0.9719  \\
        $\ket{\Psi^-} = \frac{1}{\sqrt{2}}(\ket{\tilde{01}} - \ket{\tilde{10}})$ & 0.9853 & 0.9782 \\
        \hline
        \textbf{Average} & 0.9849  & 0.9743 \\
        \hline
    \end{tabular}
    \caption{Fidelity $\mathcal{F}$ and concurrence $\mathcal{C}$ for each Bell state and average fidelity at temperature $T=0.4$ K and $T_1^{(1)} = T_1^{(2)} = 150$ ns.}
    \label{tab:bell_states}
\end{table}

\section{Discussions}

Our results highlight the intrinsic limits of achieving high-fidelity quantum gate operations in static exchange coupled qubits. In the following, we will discuss (i) how the spectrum of optimized pulses is reshaped by decoherence and (ii) the limits on optimized fidelity with realistic parameters.

\subsection{Spectrum of optimized pulses}

{\color{black}A central practical question is how QOCT improves fidelity. This insight should guide future experimental pulse designs. To build this intuition, we analyze the control fields in the frequency domain and track how decoherence reshapes the spectral content of the Krotov-optimized pulses.} Figure~\ref{fig:spectrum_vs_relaxation} summarizes how energy relaxation and pure dephasing reshapes the spectrum of the Krotov-optimized control. Throughout, we quantify the \emph{width} $w$ by the full width at half maximum (FWHM) of each spectral line and the \emph{height} $h$ of the corresponding peak amplitude. As in the closed-system case discussed above, the optimal spectrum indicates that Krotov method exploits the full set of available control degrees of freedom; this behavior persists even with relaxation. We do not consider the long-coherence-time limit here, because the optimization setup differs fundamentally from the closed-system case. Instead of aTrotterized exponential unitary propagator, we evolve density matrices using the Lindblad master equation, employing three encoded density matrices to reduce computational cost~\cite{Goerz2014}. While a finite coherence time in the Markovian Lindblad formalism will induce a decrease in fidelity regardless of the shape of the applied pulse~\cite{Jankovi2024}, by acting on the pulse-/Hamilonian-dependence of the average gate fidelity~\cite{hartmann2025nonlinearity}, an optimized pulse can mitigate this decrease as illustrated in the Supplemental Material Fig.~\ref{fig:optimal_works}.

Across both channels, the optimized pulses remain resonant as there is no discernible peak drift from the transition frequencies, while all the peak envelopes broaden with larger error rates. In particular, for energy relaxation, the amplitude of the peaks appear to remain more or less constant, while for pure dephasing, the peak height increases as dephasing strengthens, as illustrated in the lower-right panels of Figure~\ref{fig:spectrum_vs_relaxation}(a) and (b). The increase in bandwidth seems to increase monotonically, linearly even, with $T_1$ (resp. $T_\phi^{-1}$) for energy relaxation (resp. pure dephasing), as illustrated in the lower-left panels of Figure~\ref{fig:spectrum_vs_relaxation}(a) and (b). Dephasing broadens the system’s susceptibility~\cite{sobel2012excitation}, one can expect a broader control spectrum to overlap that broadened response better and thus be more robust to phase randomization and energy relaxation. We would like to further note that due to the finite pulse duration, the Fourier transform of the pulse has a minimum width on the of order $1/\tau$, which explains the coarse appearance of the spectrum profile.

Physically, amplitude increase concentrates more energy near resonance, while broadened spectral lobes could indicate, by Fourier duality, faster time-domain features. In the Supplemental Material Fig.~\ref{fig:spectrumdifferentlambda}, a saturation in pulse width is observed for the energy relaxation channel with a Krotov functional weighted differently; this could be explained by the running cost in the Krotov functional (see Eq.~(\ref{eq::running_cost})), which can act as a soft amplitude cap: once the marginal fidelity gain from making the pulse stronger or more jagged equals the marginal penalty from the running cost, the gradient balances and the update stalls.

\begin{figure*}[t]
    \centering
    \includegraphics[width=\linewidth]{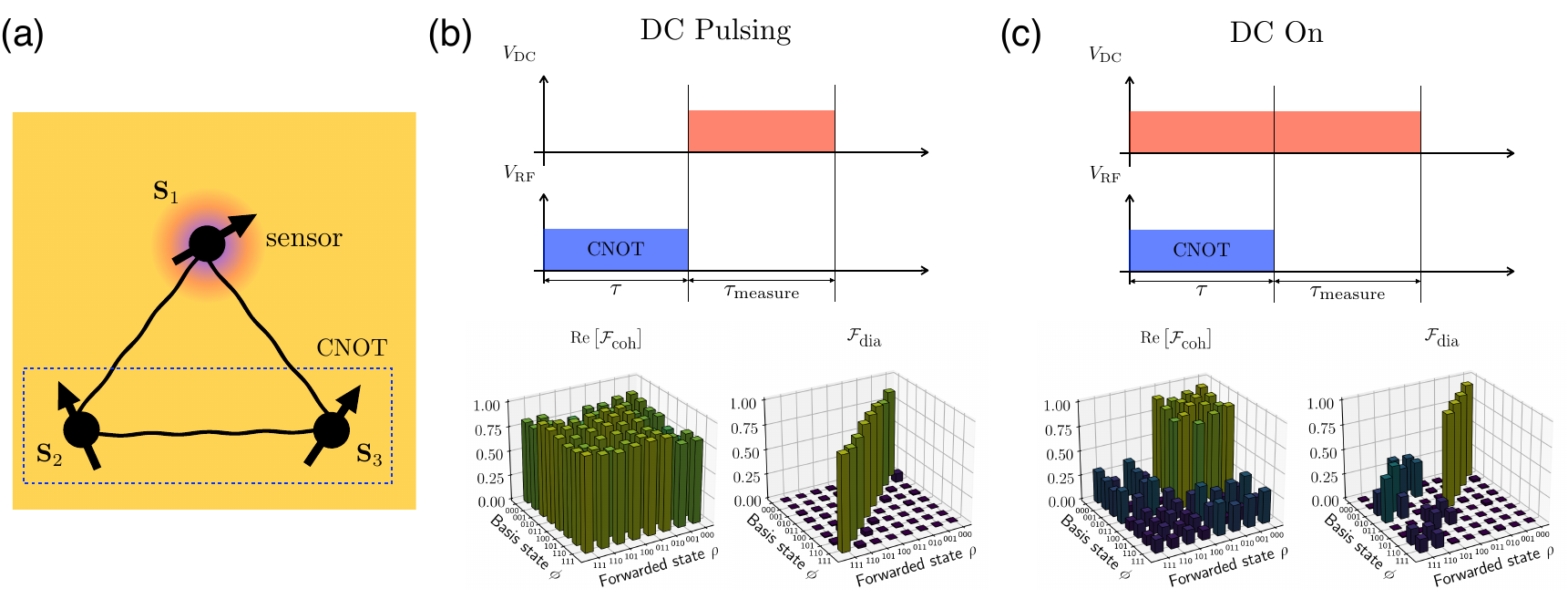}
    \caption{\textbf{Performance of QOCT with realistic parameters.}  
    (a) CNOT operation is implemented on two remote qubits. (b) Coherent and diagonal fidelity with using DC pulsing scheme, such that lifetimes are $T_1^{(1)} = T_1^{(2)} = T_1^{(3)} = 1000$ ns. The average gate fidelity is $\mathcal{F}_\text{avg} = 0.887$. (c) Coherent and diagonal fidelity without using DC pulsing scheme, $T_1^{(1)} = 10$ ns and $T_1^{(2)} = T_1^{(3)} = 1000$ ns. The average gate fidelity is  $\mathcal{F}_\text{avg} = 0.446$.
    The parameters of the drift Hamiltonian are $\omega_1 = 17 \text{ GHz}, \; \omega_2 = 15 \text{ GHz}, \; \omega_3 = 13 \text{ GHz}$, $J_{12} = 0.1 \text{ GHz}, J_{23} = 0.35 \text { GHz}, J_{13} = 0.6 \text{ GHz}$. {\color{black} Temperature is $T = 0.1 \text{ K}$} and the pulse width is $\tau = 12.5$ ns. See the definition of basis state and forwarded state in the main text below Eq.~\eqref{eq:average_gate_fidelity}.}
    \label{fig:specific_case}
\end{figure*}

These trends suggest clear strategies for improving fidelity at fixed gate duration: allowing slightly higher spectral width or peak amplitude (subject to hardware constraints) can offset decoherence effects, particularly dephasing. Conversely, if spectral width is the limiting factor, shortening the gate duration or reweighting the relevant penalty in the cost functional may prevent the observed saturation and recover fidelity without excessive spectral spreading.
When the above is taken into account, high-fidelity gate operations can be achieved in the experiments.

\subsection{Limit with realistic parameters}
\label{sect:realistic_parameters}

We will now discuss the results of our optimized quantum control simulations in the context of the experimentally available parameter range. 
We focus on the most studied platform of Ti atoms deposited on ultrathin layers of MgO grown on a silver substrate. Currently, Ti atoms on 2 monolayers (ML) and 3 ML of MgO are available according to experimental results. The spin lifetime of Ti has been studied on 2 ML of MgO and lies around $T_1=150$ ns (for the remote qubits); for 3 ML Ti has been studied by ESR-STM~\cite{phark2025spinstateengineeringsingletitanium}, but the lifetime has not been established. Based on the difference in lifetimes for Fe on 2 and 3 ML of MgO however~\cite{Paul2017}, we can expect at least one order of magnitude increase in $T_1$. We therefore set $T_1\leq 1000$ ns as a conservative estimate for 3 ML of MgO. 
The most commonly reported parameters are summarized in Tab.~\ref{tab:decoherence_time}.

\begin{table}[h!]
    \renewcommand{\arraystretch}{1.5}
    \centering
    \begin{tabular}{|c|c|c|c|c|}
        \hline
        Substrate & $T_1^{\mathrm{Sensor}}$ (ns) &$T_1^{\mathrm{Remote}}$ (ns)  & $T$ (K) & Refs. \\
            \hline
        2 ML & $5< T_1 < 10$  & $100< T_1 < 150$  & 0.4 & ~\cite{Phark2023DoubleResonance, Wang2023science, Wang2023universal}\\
        \hline
       3 ML  & $5< T_1 < 10$ & $T_1 \geq 1000$ & 0.4 & \cite{Paul2017, phark2025spinstateengineeringsingletitanium} \\
       \hline
    \end{tabular}
    \caption{Experimentally reported lifetime $T_1$ times of Ti sensor and remote spins on MgO/Ag(100). ML stands for monolayers.}
    \label{tab:decoherence_time}
\end{table}

Here, we parametrized the drift Hamiltonian using the Larmor frequencies and upper/lower bounds of coupling strengths from Ref.~\cite{Wang2023science}. The coupling strength between remote qubits is chosen so that resonant frequencies are well separated. To explore the limits of this platform, we set the temperature $T = 0.1 \text{ K}$, which is achievable in a dilution fridge and consider energy relaxation as the relevant decoherence mechanism. We assume that Ti atoms on 3 ML can achieve at least $T_1^{(2)} = 1000$ ns and set the coherence time of sensor qubit to $T_1^{(1)} =10$~ns. To account for realistic driving strengths the  pulse width is set to $\tau = 12.5
\text{ ns}$, which is the shortest possible pulse width (strongest driving) realized in experiments. We consider two scenarios for the quantum-coherent manipulation. First, we used a \textit{DC pulsing scheme} where the DC bias of the sensor spin is only active after the quantum control during the readout phase (see Fig.~\ref{fig:specific_case}(b)). In this case the sensor and remote spins are subject to the same relaxation mechanism governed by the spin-substrate energy relaxation and we set $T_1^{(1)}=T_1^{(2)}=T_1^{(3)}=1000$ ns, which corresponds to the lowest energy relaxation rates as shown in Fig.~\ref{fig:asymmetry_fidelity}(b). In such a case, one can expect a high gate fidelity as the coherence time is much longer than the pulse width. This corresponds to the protocol suggested in previous theoretical works~\cite{Broekhoven2024protocol}, but is not the one used in recent experiments. Instead, Refs~\cite{Wang2023science,Wang2023universal, Phark2023DoubleResonance} used a \textit{DC always-on} scheme where the DC bias is on during quantum gate operations and readout phase (mostly due to technical reasons related to drift compensation), as illustrated in Fig.~\ref{fig:specific_case}(c). In such a scenario, we investigate the realization of CNOT gate on two remote qubits as illustrated in Fig.~\ref{fig:specific_case}(a). Using Krotov method, the CNOT gate fidelity is improved from 0.446 to 0.887 when using the DC pulsing scheme. The result suggests that the DC pulsing scheme is highly beneficial to achieve high-fidelity quantum gates.
Notice that the spectrum of optimized pulse here also shows four pronounced peaks at the resonant frequencies, similar to the closed-system case (see Supplemental Material Fig.~\ref{fig:closed_CNOT}).

The distribution of coherent and diagonal fidelities over the Liouvillian basis provides a further insight into the performance of quantum gates, compared to a mere average gate fidelity. Therefore we further analyze the components of the average gate fidelity. The average gate fildelity can be written in the following form~\cite{Pedersen2007fidelity}:
 \begin{equation}
\begin{aligned}
\mathcal{F}_{\text{avg}} &= \frac{1}{N (N+1)} \left( \mathcal{F}_\text{coh} + \mathcal{F}_\text{dia} \right), \\
\mathcal{F}_\text{coh} &= \sum_{i,j=1}^N  \big\langle
\phi_i \big\vert U_\text{target}^\dagger \; \rho_{ij} \; U_\text{target} \big\vert \phi_j \big\rangle, \\
\mathcal{F}_\text{dia} &= \sum_{i,j=1}^N  \big\langle
\phi_i \big\vert U_\text{target}^\dagger \; \rho_{jj} \; U_\text{target} \big\vert \phi_i \big\rangle.
\end{aligned}
\label{eq:average_gate_fidelity}
\end{equation}
where $\ket{\phi_i}$ denotes the $i$-th \textit{basis state}, and  
$\rho_{ij} = \mathcal{E}\!\left( \ket{\phi_i}\!\bra{\phi_j} \right)$ is the output obtained by applying the dynamical map $\mathcal{E}$ to the Liouvillian basis operator $\ket{\phi_i}\!\bra{\phi_j}$ (we refer this one as \textit{forwarded state}).  
$U^\dagger_\text{target}$ is the desired unitary transformation and $N$ denotes the dimension of the Hilbert space of the drift Hamiltonian. It is worth to break down the average fidelity into two parts, as shown in Eq.~\eqref{eq:average_gate_fidelity}: $\mathcal{F}_\text{coh}$ quantifies the preservation of coherence, while
$\mathcal{F}_\text{dia}$ quantifies the
population fidelity. Coherent fidelity can take complex values, yet only the real part is relevant as the imaginary part will vanish when summing over all basis states. 

As shown in Fig.~\ref{fig:specific_case}(c), although the average fidelity is 0.446, the fidelities associated with individual state transitions exhibit noticeable variation. This observation indicates that optimizing each transition separately within the state-to-state approach can yield higher fidelities compared to the gate optimization.

With the DC pulsing scheme and thus long coherence times for all qubits, both coherent and diagonal fidelities are high and nearly uniform across the Liouvillian basis. However, without the DC pulsing scheme and thus short coherence time for the sensor qubit, the coherent and diagonal fidelities are significantly reduced and skewed towards low-energy states $\ket{000}$ and $\ket{\tilde{001}}$. This is due to energy relaxation where higher-energy states decay into lower-energy ones. With short coherence time of sensor qubit $T_1^{(1)} = 10$ ns, Krotov method can only perform well for low-energy states, while the fidelity for high-energy states is significantly reduced. In addition, the distribution of diagonal fidelity is asymmetric. This feature is a result of the unidirectional energy relaxation.

\subsection{\color{black}Strategies to improve fidelities in this platform}

{\color{black}In this work, we investigated how quantum optimal control theory (QOCT), implemented through Krotov-based optimization, interacts with the dynamical constraints of static exchange-coupled spin qubits. Closed-system analyses serve as useful reference points, where QOCT reaches gate fidelities close to $\mathcal{F}=100\%$ almost independent of the drift Hamiltonian parameters. The more relevant regime for surface-based qubits is one in which asymmetrical relaxation and dephasing are explicitly incorporated into the control landscape. When these open-system effects are included, the optimized control fields exhibit characteristic spectral changes that depend on the dominant decoherence mechanism. We observe consistent improvement of fidelity and concurrence with QOCT compared to un-optimized or Rabi-synchronized driving. For coherence parameters that are realistic for current experiments (remote-qubit lifetimes exceeding $1~\mu\mathrm{s}$ and weak pure dephasing), fidelities above $90\%$ appear feasible using experimentally accessible pulse shapes and DC bias pulsing. While extending noise-aware QOCT to larger systems increases computational cost and pulse complexity, the essential advantage persists when the optimization incorporates the decoherence channels that dominate the platform rather than relying on closed-system targets. In this light, the fidelity gains obtained under experimentally motivated pulsing schemes should be viewed as indications of how control strategies benefit from aligning pulse design with open-system dynamics. Future investigations into scalability, and robustness against parameter fluctuations may clarify how these methods extend beyond the present setting, and where the balance between model completeness and practical control performance lies for future implementations.

Finally, we remark that based on our results, improvements in the gate  performance of this platform can be achieved along the following trajectories: First, enhancing the $T_1$ time of the remote qubits (Fig.~\ref{fig:asymmetry_fidelity}(b)), which can be achieved by better decoupling between the qubits and the metallic substrate~\cite{phark2025spinstateengineeringsingletitanium}. Second, increasing the driving strength thereby reducing the gate time $\tau$ with respect to the lifetime will reduce the influence of noise during the gate operation. This can be achieved experimentally by optimizing the local driving field~\cite{Reina2025driving}. Third, lowering the temperature to maintain a high state polarization for a given splitting (Fig.~\ref{fig:qoct_entanglement_open}(c)). Combined with the DC pulsing scheme presented here, gate-fidelities exceeding $\mathcal{F} \gtrsim0.9$ should be achievable straightforwardly.
}

%\pagebreak
%\newpage
%\clearpage

\section{Methods}

\textbf{Numerical simulations} All numerical simulations were carried out using the \textsc{QuTiP}~\cite{JOHANSSON2012qutip} package (Version~4.7.6) together with the \textsc{Krotov}~\cite{Krotov2019} Python package (Version~1.3.0). These packages provide efficient tools for quantum dynamics simulations and gradient-based optimal control. 

\textbf{Computation of time evolution operators of closed systems} We solve the time evolution using \textit{Trotterization} for closed quantum systems. Suppose duration for the control pulse is $\tau$, we divide the total time into $N_\text{seg}$ segments, each with duration $\Delta t = \tau / N_\text{seg}$. The time-evolution operator for each segment is approximated as
\begin{equation}
    U(t_{n+1}) = \exp \left[ -i H(t_n) \Delta t \right] U(t_n),
\end{equation}
where $U(0) = I$ is the initial unitary operator and $H(t_n)$ is the Hamiltonian at time $t_n \equiv n \Delta t$. The full time-evolution operator is then given by the product of all segments
\begin{equation}
    U(\tau) = \prod_{n=0}^{N_\text{seg}-1} \exp \left[ -i H(t_n) \Delta t \right].
\end{equation}
The approximation becomes exact in the limit of $N_\text{seg} \rightarrow \infty$. In our simulations, we choose $N_\text{seg}$ such that $\Delta t = 0.01~\text{ns}$ and the results converge.

\textbf{Computation of spin dynamics of closed systems} 
Suppose the initial density matrix of the two-qubit system is $\rho(0)$, the density matrix at time $t$ is given by
\begin{equation}
    \rho(t) = U(t) \rho(0) U^\dagger (t),
\end{equation}
with $U(t)$ being the time-evolution operator. The spin of second qubit at time $t$ is computed as
\begin{equation}
    \langle S_2^{\alpha} (t) \rangle = \text{Tr}_1 \left[ S^\alpha \rho(t) \right],
\end{equation}
where $\alpha = x,y,z$ and $\text{Tr}_1$ is the partial trace over the first qubit. In the interaction picture, one needs to perform an additional transformation to the density matrix as 
\begin{equation}
    \rho_I (t) = e^{i H_0 t} \rho(t) e^{-i H_0 t},
\end{equation}
where $H_0$ is the drift Hamiltonian.

\textbf{Computation of gate fidelity of closed systems} In the lab frame, the evolution that corresponds to drift Hamiltonian should be taken into account. As for the NOT gate on the second qubit, the target unitary operator is given by~\cite{nielsen2002} 
\begin{equation}
\begin{aligned}
U_\text{target}(t) = \exp(- i E_1 t) \ket{00} \bra{\tilde{01}} + \exp(-i E_2 t) \ket{\tilde{01}}\bra{00} \\+ \exp(-i E_3 t) \ket{\tilde{10}} \bra{11} + \exp(-i E_4 t) \ket{11}\bra{\tilde{10}}.
\end{aligned}
\end{equation}
The fidelity of the gate operation is then computed using
\begin{equation}
    \mathcal{F}_{U(t), U_\text{target}} = \frac{\vert \text{Tr} [ U(t)^\dagger U_\text{target}(t) ] \vert^2 + d }{d(d+1)},
    \label{eq:closed_fidelity}
\end{equation}
where $d$ is the dimension of the Hilbert space (for example, $d=4$ for a two-qubit system).

\textbf{Quantum optimal control theory with Krotov method}
Quantum optimal control theory (QOCT) is a set of methods to devise external control fields that drive the system toward a desired objective. Assuming the system is described by the following time-dependent Hamiltonian
\begin{equation}
    \hat{H}(t) = \hat{H}_0 + \sum_j u_j(t) \hat{H}_j,
\end{equation}
where $\hat{H}_0$ is the time-independent drift Hamiltonian and $u_j (t)$ are time-dependent control functions. The goal is to solve the set of control fields, defined over a duration $\tau$, to achieve a specific purpose, such as implementing a desired unitary operation (i.e. quantum gate) or transition from an initial to a target state. The former is called \textit{gate optimization} and the latter is \textit{state-to-state optimization}. The gate optimization is essential for realizing universal quantum gates where the operation must be correct for any of the basis states, while the state-to-state optimization targets a specific transition from an initial to a target state over the control time. We particularly employ Krotov method for our work since it is well-suited for both closed and open quantum systems, and is efficiently implemented in the Python Krotov library~\cite{Krotov2019}. 

A minimal introduction to Krotov method for our two-qubit system in Eq.~\eqref{eq:two_qubit_hamiltonian} with single control field can be presented as follows. The Krotov method can be formulated as an optimization process to minimize the following cost functional form 
\begin{equation}
J\left[ \vert \phi^i_k (t) \rangle, u^i (t) \right] = J_T (\vert \phi^i_k (\tau) \rangle) + \int_0^\tau g_a (u^i (t)) dt,
\label{eq:krotov_functional}
\end{equation}    
where $\vert \phi_k^i (t)\rangle$ are the time-evolved the initial states $\vert \phi_k^i (0) \rangle$ under the control field $u^i (t)$ at iteration $i$. (Indeed, there exists an additional term to penalize state population in a subspace, yet we don't employ it in our work.) For state-to-state optimization $k=1$, while for gate optimization $k=1,2,3,4$ for the four basis states $\ket{00}, \ket{\tilde{01}}, \ket{\tilde{10}}, \ket{11}$. The first term is the main functional cost, which essentially measures the distance between the desired unitary transformation and actual evolution. Given a set of initial states $\vert \phi_k \rangle$ to evolve to a set of target states $\vert \phi_k^\text{target} \rangle$, their distance can be quantified by the fidelity
\begin{equation}
    f_k = \bra{\phi_k^\text{target}} \phi_k (\tau) \rangle,
    \label{eq:f_k}
\end{equation}
and the function $J_T$ reads 
\begin{equation}
    J_T = 1 - \frac{1}{N} \text{Re} \left[ \sum_{k=1}^N f_k \right].
    \label{eq:J_T}
\end{equation}

The second term is the running cost functional, which penalizes the deviation of the control field from a reference field $u_\text{ref} (t)$, and is defined as
\begin{equation}
    \label{eq::running_cost}
\begin{aligned}
    g_a (u^i (t)) &= \frac{\lambda_a}{S (t)} \left[ u^i (t) - u^i_\text{ref} (t) \right]^2; \\
    \quad u^i_\text{ref} (t) &= u^{i-1} (t),
\end{aligned}
\end{equation}
with $\lambda_a$ being the inverse step width and $S (t)$ is an ``update shape" function that enforces smooth on/off switches. In our simulations, we choose $u^0_\text{ref} (t)$ as a flattop pulse with smooth turn-on and turn-off edges, and $S (t)$ is also a flattop pulse~\cite{Krotov2019}. The choice of $\lambda_a$ is determined at the beginning of the optimization process, and is kept constant throughout the iterations.

The above description works for the closed quantum system, i.e. without decoherence. In the presence of decoherence, we adopt the method in Ref.~\cite{Goerz2014} to incorporate the Lindblad master equation and the three representative states of Liouville space. In this case, the cost functional is modified as
\begin{equation}
    J_T = 1 - \sum_{i=1}^{3} \frac{w_i}{\text{Tr} 
    \left[ \hat{\varrho}(0) \right]} 
    \text{Re} \left\{ \text{Tr} 
    \left[ \hat{U}_\text{target} \hat{\varrho}_i(0) \hat{U}^\dagger_\text{target} \hat{\varrho}_i (T) \right] \right\},
\end{equation}
where $\hat{\varrho}_i \; (i = 1,2, 3)$ denote three encoded density matrices of orthonormal basis of Liouville space, and $w_i$ are the corresponding weights satisfying normalization condition. The choice of $w_i$ is also determined at the beginning of the optimization process, and is kept constant throughout the iterations. The dimension of Liouville space is $2^{2N}$ for $N$ qubits.

\textbf{Optimization Parameters} The control optimization was performed using Krotov method, which ensures monotonic convergence of the cost functional. The time evolution was discretized into $N_t$ time steps with $\Delta t = \tau / N_t$, where $\tau$ is the total propagation time. {\color{black} Throughout this work, we set $\Delta t = 0.01$ ns, which is approximately two-times smaller than the smallest time scale in the system (i.e., the inverse of the largest Larmor frequency of the Hamiltonian), to ensure numerical convergence. Influence of undersampling/oversampling time segment $\Delta t$ on the optimized fidelity is discussed in the Supplemental Material Fig.~\ref{fig:sm_figure15}.}
Propagation was performed using the  exponential matrix propagator \texttt{expm} for closed-system dynamics and the built-in ordinary differential equations solver \texttt{DensityMatrixODEPropagator} for open-system calculations, both implemented within the Krotov package. A flattop update shape function $S(t)$ was employed to enforce smooth boundary conditions on the control fields. The step-size parameter $\lambda_a$ was chosen separately for each optimization to balance the update strength with numerical stability. In particular, we used $\lambda_a = 10$ for NOT gate optimization and $\lambda_a = 0.01$ for Bell-state preparation.

\textbf{Convergence criteria} 
% The convergence of the optimization was monitored by tracking the optimization functional and the corresponding fidelities as a function of the iteration number.
{\color{black} The convergence can be monitored through changes in the optimization functional $J_T$, which depends on the fidelity as defined in Eqs.~\eqref{eq:f_k} and~\eqref{eq:J_T} and corresponds to the infidelity. In closed-system optimizations, unit fidelity is known to be attainable and we define convergence as $J_T\leq 10^{-6}$. Alternatively, convergence can be conditioned on the change in $J_T$, $\Delta J_T=|J_T^N-J_T^{N-1}| \leq 10^{-6}$ between successive iterations which means that fidelity became stationary. For open-system dynamics, the maximum achievable fidelity is lower than 1 and it is hard to predict a reliable target fidelity. Instead, we found it more stable to run the optimization for a fixed number of iterations, $N^{\text{max}}_\text{iter}=600$ for closed systems and $N^{\text{max}}_\text{iter}=100$ for open systems, and verified that the infidelity remained stationary ($\Delta J_T \leq 10^{-6}$) over a significant number of iterations, which was the case in all simulations. This way we ensured a balance between computational time and optimization outcome. }
% In some rare cases neither fidelity nor the stationary criterion can be achieved. In such cases, we instead stop the optimization at a fixed number of iterations $N_\text{iter}$.}

%The best number of $N_\text{iter}$ was chosen by identifying the point at which the fidelity becomes stationary within a tolerance no larger than $10^{-3}$ on the gradient ($10^{-6}$ in the case of closed systems). 
%This procedure ensured that the reported results correspond to the optimal balance between computational efficiency and control performance. 

\textbf{Concurrence} 
Given a two-qubit density matrix $\rho$, the concurrence is defined as~\cite{wootters1998entanglement}
\begin{equation}
    C(\rho) = \max \left\{ 0, \, \lambda_1 - \lambda_2 - \lambda_3 - \lambda_4 \right\},
\end{equation}
where the \(\lambda_i\) are the eigenvalues, arranged in decreasing order, of the following Hermitian matrix
\begin{equation}
    R = \sqrt{ \sqrt{\rho} \, \tilde{\rho} \, \sqrt{\rho} },
\end{equation}
where \begin{equation}
    \tilde{\rho} \equiv (\sigma_y' \otimes \sigma_y')\, \rho^* \, (\sigma_y' \otimes \sigma_y'),
\end{equation}
with $\rho^*$ the complex conjugate of $\rho$.
Here $\sigma_y'$ denotes the Pauli-$y$ matrix in eigenstate basis, with
\begin{equation}
\sigma_y' \otimes \sigma_y' = -\ket{00}\bra{11} + \ket{\tilde{01}}\bra{\tilde{10}} + \ket{\tilde{10}}\bra{\tilde{01}} - \ket{11}\bra{00}.
\end{equation}

\textbf{Collapse operators}
We used collapse operators acting on the eigenstates of the system as defined in Ref.~\cite{Phark2023DoubleResonance}. Description for collapse operators can be found in the Supplemental Material.

\section{Acknowledgments}
All authors acknowledge support from the Institute for Basic Science under grant IBS-R027-D1. We acknowledge fruitful discussions with Nicol\'as Lorente, Yaowu Liu, and {\color{black}Dan-Chi Nguyen} during the writing of this manuscript. We thank Yaowu Liu and J.-G. Hartmann for carefully reading our manuscript.

\section*{Data Availability}
% The data supporting the findings of this study are available from the corresponding author upon reasonable request.
The datasets generated during the current study are not publicly available because they require  analysis scripts that are not yet curated for public release, but are available from the corresponding author on reasonable request.

\section*{Author Contribution}
%CW conceived the paper, HAL, ST and DJ performed numerical simulations. All authors contributed to the discussion and the writing of the manuscript.

C.W. and H.-A.L. conceived the ideas of the manuscript, H.-A.L., S.T., and D.J. performed the numerical simulations. All authors contributed in writing the manuscript under the supervision of C.W.

\section*{Competing interests}
The authors declare no competing interests.

\bibliography{ref.bib}

\newpage

\clearpage
\onecolumngrid

\begin{center}
{\Large \centering \textbf{Supplemental Material \\ Overcoming limitations on gate fidelity in noisy static exchange-coupled surface qubits}}
\end{center}

\renewcommand{\thesection}{S\arabic{section}}
\renewcommand{\thetable}{S\arabic{table}}
\renewcommand{\thefigure}{S\arabic{figure}}
\renewcommand{\theequation}{S\arabic{equation}}
\setcounter{figure}{0}
\setcounter{table}{0}
\setcounter{section}{0}
\setcounter{equation}{0}

\section{Choice of Zeeman-like basis states}

For most of the simulations we used a drift Hamiltonian with $\omega_1 = 20 \pi \text{ GHz}, \omega_2 = 14 \pi \text{ GHz}, J = 5 \text{ GHz}$. This Hamiltonian is partially optimized to result in well-separated states as well as being compatible with the most widely available experimental setups in terms of the transmission window of the RF signals. The eigenstates of this Hamiltonian are not perfect Zeeman product states ${\ket{00}, \ket{01}, \ket{10}, \ket{11}}$ however their hybridization is low. The two hybridized eigenstates are 
\begin{equation}
    \ket{\tilde{01}} \approx 0.992 \ket{01} - 0.129  \ket{10}, \quad
    \ket{\tilde{10}} \approx 0.129 \ket{01} + 0.992 \ket{10}.
\end{equation}
As long as the system eigenstates are close to Zeeman product states $\ket{01}$ and $\ket{10}$, we could model the energy relaxation and dephasing of each spin independently.

\section{Gate fidelity with phase difference}

\begin{figure}[h!]
    \centering
    \includegraphics[width=0.7\linewidth]{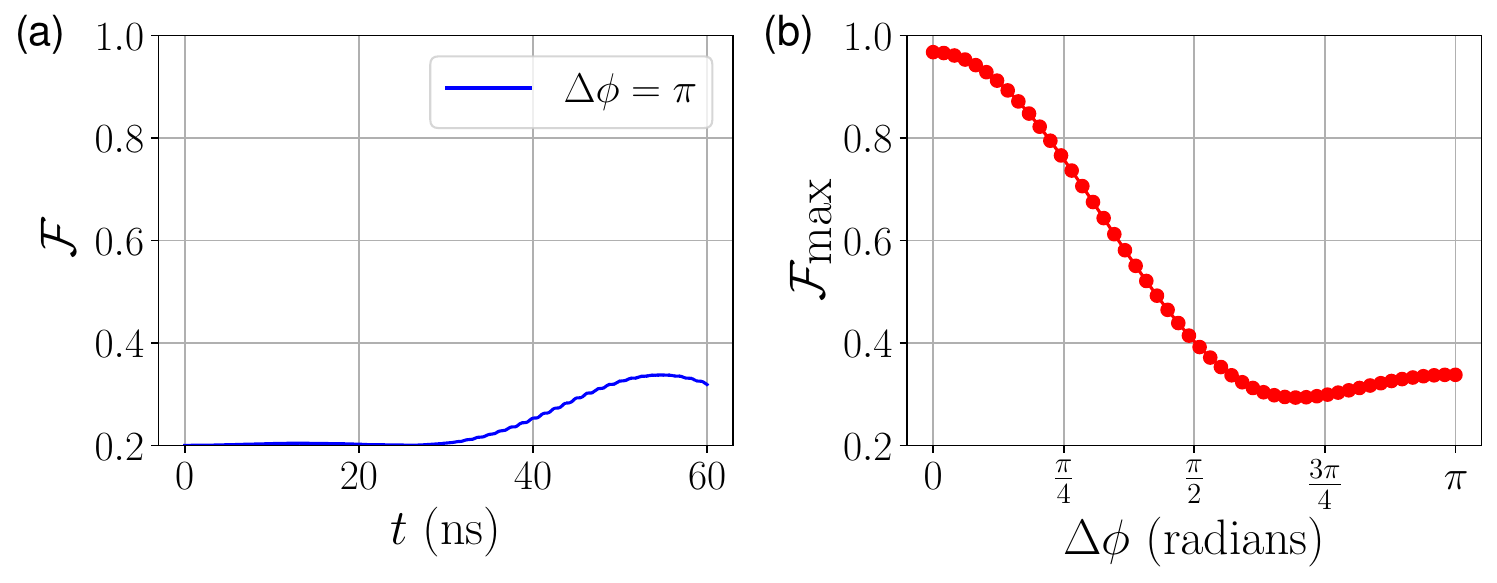}
    \caption{(a) Fidelity of NOT gate as a function of time, with phase difference $\Delta \phi = \phi_2 - \phi_1 = \pi$. (b) Maximal fidelity as a function of phase difference $\Delta \phi$. The drift Hamiltonian parameters are $\omega_1 = 20\pi~\text{GHz}$, $\omega_2 = 14\pi~\text{GHz}$, and $J = 5~\text{GHz}$. Amplitudes of two resonant frequencies are $\Omega_1 = \Omega_2 = \pi/25$.}
    \label{fig:angle}
\end{figure}

We revisit the realization of NOT gate for two-qubit system. In the main text, we have set two resonant frequencies with identical phase, $\phi_1 = \phi_2 = 0$, as shown in Eq.~\eqref{eq:driving_pulse}. We examine the effect that the two frequencies have a phase difference, where the bichromatic driving pulse can be generally written as
\begin{equation}
\mathcal{D}(t) \equiv \Omega_1 \cos (\omega_{\text{RF1}}t +\phi_1) + \Omega_2 \cos (\omega_{\text{RF2}}t +\phi_2).
\end{equation}
The phase difference is defined as 
\begin{equation}
    \Delta \phi = \phi_2 - \phi_1.
\end{equation}
The effect of phase difference is summarized in Fig.~\ref{fig:angle}, where zero phase difference gives the highest fidelity for the NOT gate on second qubit.

\newpage
\section{Krotov method for closed systems}

\begin{figure}[ht]
    \centering
    \includegraphics[width=0.6\linewidth]{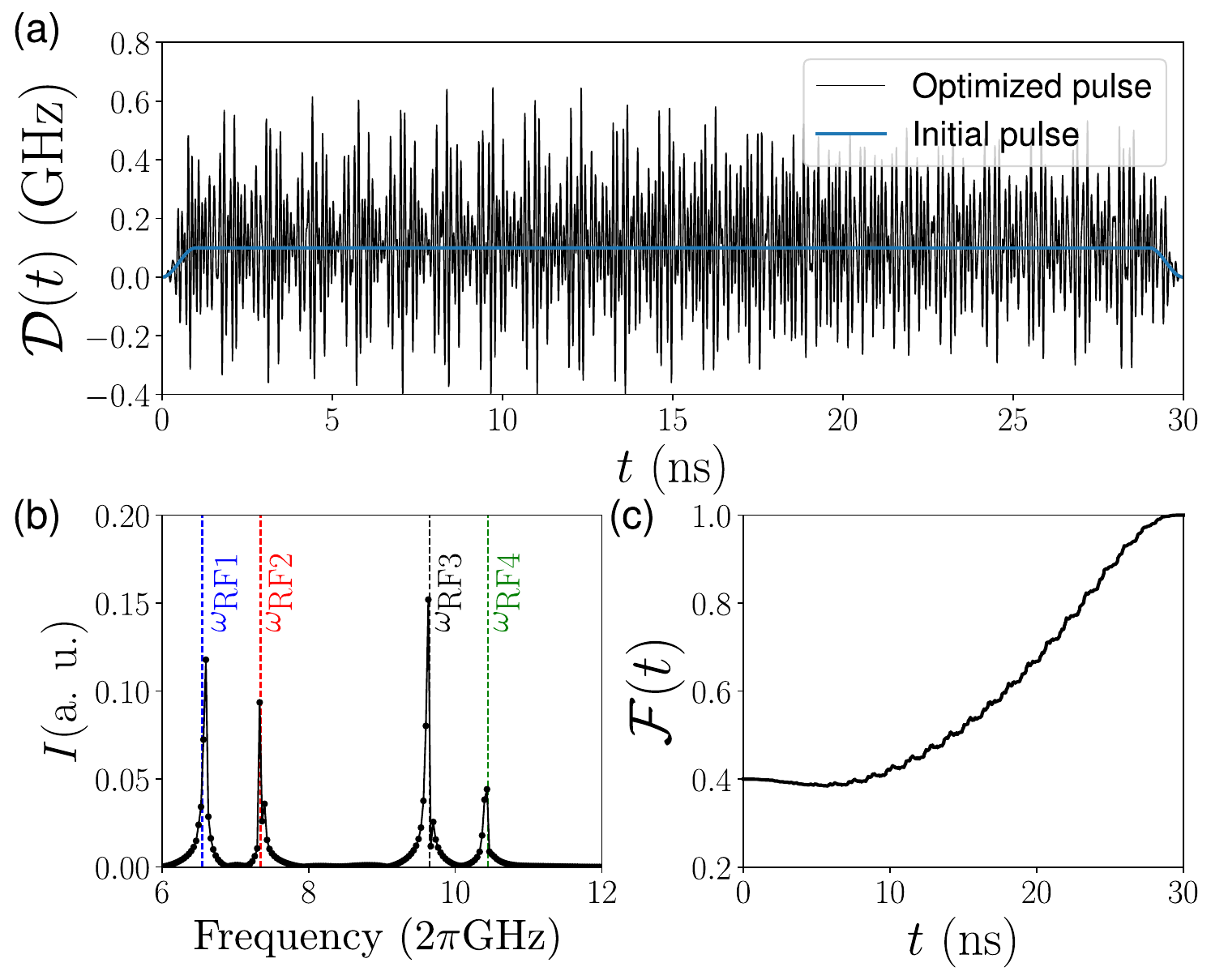}
    \caption{{Realization of a CNOT gate using Krotov method.} (a) The optimal pulse (black) compared to the initial flattop pulse (blue).
    (b) Fourier transform of the optimal pulse. (c) Fidelity of the NOT gate as a function of time. The pulse width is set to $\tau = 30$ ns. The drift Hamiltonian parameters are $\omega_1 = 20\pi~\text{GHz}$, $\omega_2 = 14\pi~\text{GHz}$, and $J = 5~\text{GHz}$. The Krotov method parameters are $\lambda_a = 0.5$ and $N_\text{iter} = 300$.}
    \label{fig:closed_CNOT}
\end{figure}

\begin{figure}[ht]
    \centering
    \includegraphics[width=0.6\linewidth]{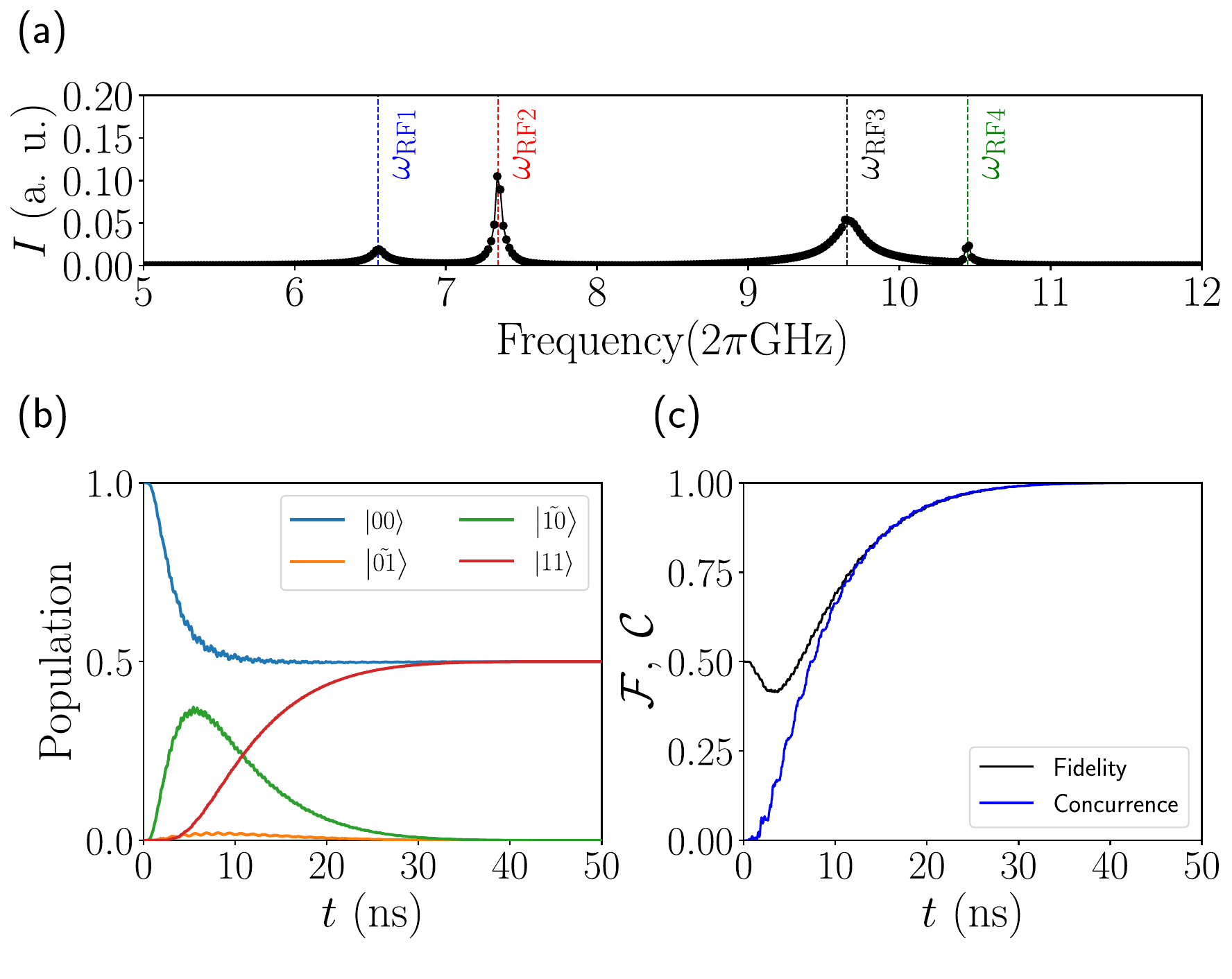}
    \caption{{Entanglement preparation in a closed system using Krotov method.} The initial state is $\ket{00}$ and the target state is the Bell state $\ket{\Phi^+} = \frac{1}{\sqrt{2}} (\ket{00} + \ket{11})$.
    (a) Fourier transformation of the optimal pulse. The pulse width is set as $\tau = 50$ ns. (b)  Population of the states as a function of time. 
    (c) Fidelity $\mathcal{F}$ between propagated state and $\ket{\Phi^+}$ and its concurrence $\mathcal{C}$ are plotted as a function of time. 
    %The fidelity is nearly 100\%  and concurrence is 1 at the end of optimal pulse.
    The parameters of drift Hamiltonian are $\omega_1 = 20 \pi \text{ GHz}, \omega_2 = 14 \pi \text{ GHz}, J = 5 \text{ GHz}$.
    }
    \label{fig:qoct_entanglement_closed}
\end{figure}

\begin{figure}[ht]
    \centering
    \includegraphics[width=0.6\linewidth]{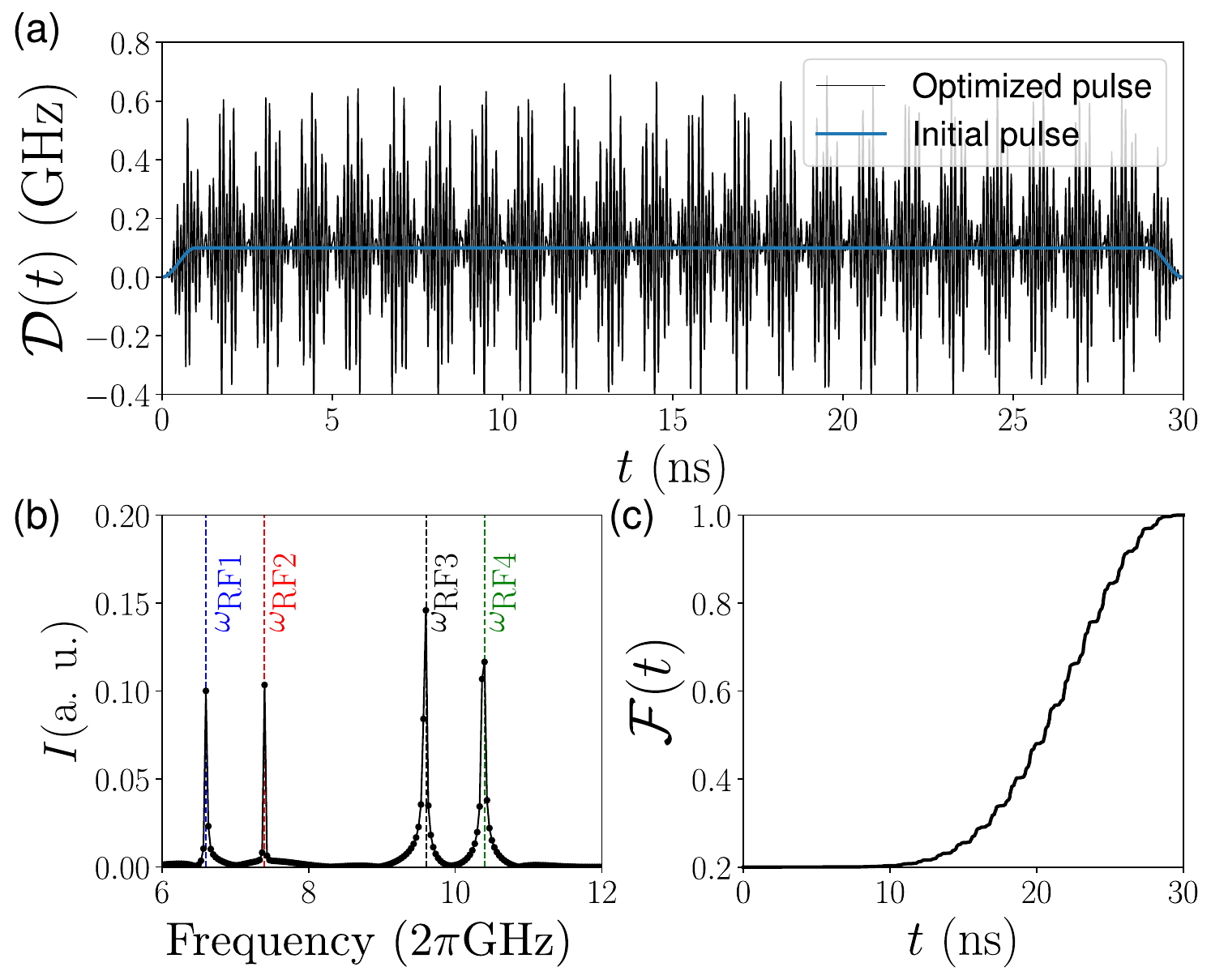}
    \caption{{Realization of a NOT gate on the second qubit with Ising drift Hamiltonian using Krotov method.} (a) The optimal pulse (black) compared to the initial flattop pulse (blue). The pulse width is set to $\tau = 30$ ns. The drift Hamiltonian parameters are $\omega_1 = 20\pi~\text{GHz}$, $\omega_2 = 14\pi~\text{GHz}$, and $J_\text{Ising} = 5~\text{GHz}$. The Krotov method parameters are $\lambda_a = 0.5$ and $N_\text{iter} = 300$.}
    \label{fig:closed_Ising}
\end{figure}

Here we present additional results of the Krotov method for closed systems with (i) CNOT gate on the second qubit, (ii) entanglement preparation, and (ii) NOT gate on the second qubit using Ising interaction. 
Case (i) only requires a monochromatic pulse~\cite{Wang2023universal}. Case (ii) requires state-to-state optimization to determine the pulse shape that steers the initial state $\ket{00}$ to the Bell state $\ket{\Phi^+} = \frac{1}{\sqrt{2}} (\ket{00} + \ket{11})$. 
%With gate-based circuit model, the entanglement preparation can be realized by a sequence of Hadamard and CNOT gates, where the corresponding pulse is a composite one that realizes the two gates in sequence.
The last case (iii) only requires a monochromatic pulse~\cite{Wang2023universal}. While the latter, whose drift Hamiltonian is given by
\begin{equation}
    H_\text{Ising} = - \omega_1 S^z_1 - \omega_2 S^z_2 + J_\text{Ising} S_1^z S_2^z,
\end{equation}
have Zeeman product states as eigenstates, thus the NOT gate on the second qubit can be realized by a bichromatic pulse with unit fidelity. The results are shown in Figs.~\ref{fig:closed_CNOT},~\ref{fig:qoct_entanglement_closed}, and~\ref{fig:closed_Ising}, respectively. The optimal pulses for all three cases show four resonant frequencies. It demonstrates that Krotov method exploits all possible degrees of freedom to achieve the target gate.

{\color{black}

% Figure~\ref{fig:J_omega} illustrates the origin of the fidelity drop observed in certain regions of the two-qubit system as a function of the coupling strength $J$ and the Larmor frequency difference $\Delta \omega = \omega_1 - \omega_2$.
The average gate fidelity $\mathcal{F}$ obtained from Krotov optimization for a closed system further does not depend strongly on the specific choice of Larmor frequency and exchange coupling, as shown in Fig.~\ref{fig:J_omega}(a).  The Krotov optimization only breaks down when the Larmor frequency difference $\Delta\omega\approx 0$, i.e. when the transitions between states of spin 1 and spin 2 become degenerate and can no longer be driven independently. This can also be understood from the character of the eigenstates of the system. For $\Delta \omega \approx 0$ the basis states $\ket{01}$ and $\ket{10}$ of the system are highly hybridized and control of individual spin flips are no longer possible. The degree of overlap can be measured by calculating the deviation $1/\sqrt{2} - \sin(\xi_1)$ from the fully entangled $1/\sqrt{2}\left( |01\rangle \pm |10\rangle\right)$ states. Here, $\xi_1 = \arctan \left(\alpha + \sqrt{\alpha^2+1} \right)$, as defined in Eq.~\eqref{eq:xi1}. Fig.~\ref{fig:J_omega}(b) shows that Krotov optimization fails exactly where $1/\sqrt{2} - \sin(\xi_1) \approx 0$ with a very sharp fall-off around $\Delta \omega \approx \pm 1$ GHz. A similar behavior occurs when $\Delta \omega = 10$ and $J<1$ GHz because $\omega_2=0$ and the splitting of the states stems only from $J$.

%degree of hybridization between the $\ket{01}$ and $\ket{10}$ states can be quantified by the analytic measure $1/\sqrt{2} - \sin(\xi_1)$, with $\xi_1$ defined in Eq.~\ref{eq:xi1}. As shown in Fig.~\ref{fig:J_omega}(b), this deviation increases both for small $J$ and for large $\Delta \omega$, demonstrating that the departure from Zeeman-like behavior depends on the interplay of these two parameters. 

\begin{figure}[ht]
    \centering
    \includegraphics[width=0.4\linewidth]{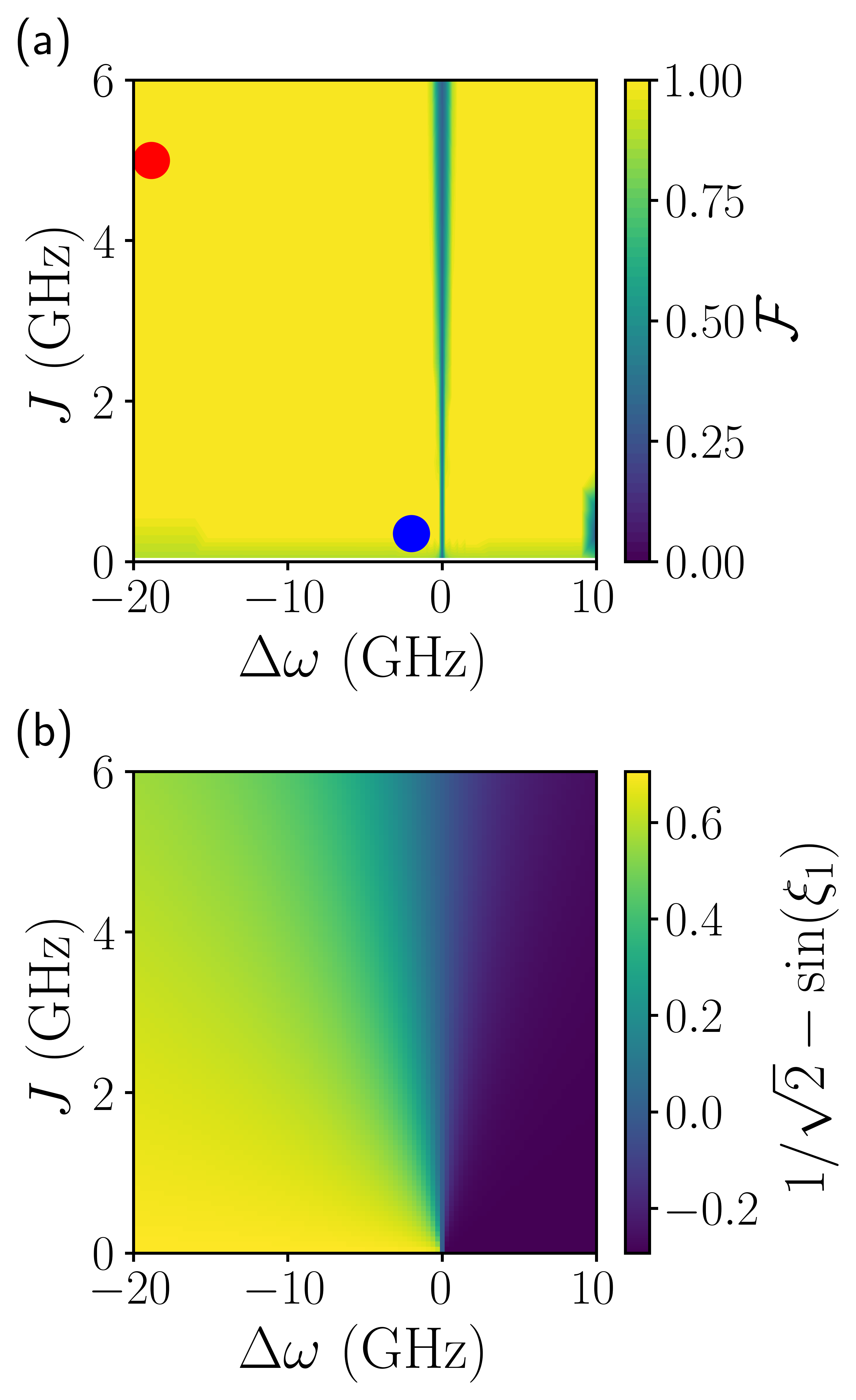}
    \caption{(a) Average fidelity $\mathcal{F}$ for a NOT gate obtained from Krotov optimization as a function of the exchange coupling $J$ and the Larmor frequency difference $\Delta \omega = \omega_1-\omega_2$ for a closed system and $ \omega_1 = 10$ GHz, and $\tau = 50$ ns. Red and blue markers indicate parameter regions studied in the main text. (b) Measure of how Zeeman-like the basis state $|\tilde{01}\rangle=-\sin{\xi_1}|01\rangle+\cos{\xi_1}|10\rangle$ of the Hamiltonian is. Shown is $1/\sqrt{2} - \sin(\xi_1)$, i.e. the deviation from the maximally entangled state. The larger $|1/\sqrt{2} - \sin(\xi_1)|$ the more Zeeman-like are the eigenstates of the system.}
    \label{fig:J_omega}
\end{figure}

% \clearpage

\clearpage

{\color{black}
\section{GRAPE method for closed systems}
\label{sec:GRAPE}
In this section, we discuss the gate fidelity optimization using GRadient Ascent Pulse Engineering (GRAPE)~\cite{Khaneja2005}, another algorithm from QOCT. Figure~\ref{fig:closed_GR} presents the result of realizing NOT gate on the second qubit for closed systems with identical drift Hamiltonian presented in the main text and Tab.~\ref{tab:wall_time} shows the comparison wall time between GRAPE and Krotov methods.

\begin{figure}[ht]
    \centering
    \includegraphics[width=0.6\linewidth]{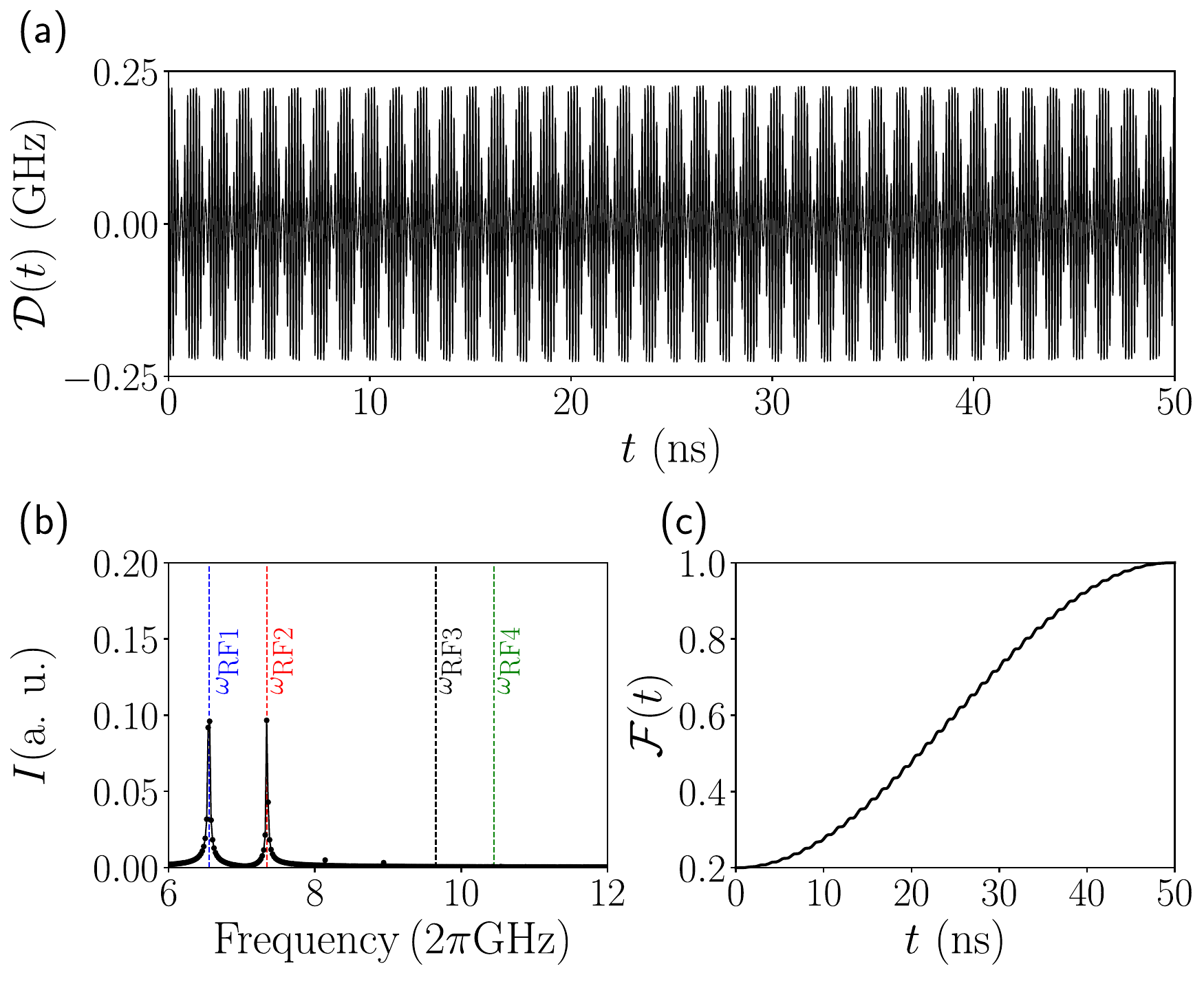}
    \caption{\textbf{Realization of a NOT gate on the second qubit using GRAPE.} (a) The optimal pulse. The pulse width is set to $\tau = 50$ ns. The drift Hamiltonian parameters are $\omega_1 = 20\pi~\text{GHz}$, $\omega_2 = 14\pi~\text{GHz}$, and $J = 5~\text{GHz}$. The optimization is initialized with a sinusoidal (SIN) pulse at frequency $\omega_{\mathrm{RF1}}$, phase $\pi/2$. (b) Fourier transform of the optimal pulse. (c) Fidelity of the NOT gate as a function of time, reaching nearly 1 at the end of the optimal pulse.}
    \label{fig:closed_GR}
\end{figure}

\begin{table}[h!]
    \renewcommand{\arraystretch}{1.5}
    \centering
    \begin{tabular}{|c|c|c|c|}
        \hline 
        \textbf{Method} & Convergence criteria & $\mathcal{F}$ & Wall time (s) \\
        \hline
        GRAPE & infidelity $ \le 10^{-6}$ & 0.9999990 & 4 \\
        \hline 
        Krotov & infidelity $ \le 10^{-6}$ & 0.9999985 & 3893 \\
        \hline
    \end{tabular}
    \caption{\textbf{Comparison of wall time between GRAPE and Krotov method.} The two methods are employed to optimized a NOT gate on the second qubit with parameters identical to those in Fig.~\ref{fig:closed_GR}. The convergence criteria for both two methods is infidelity smaller than $10^{-6}$.}
    \label{tab:wall_time}
\end{table}

The properties of optimized pulse using GRAPE are shown in Fig.~\ref{fig:closed_GR_init}. Compared with those obtained from Krotov method, their spectra exhibit peaks at non-resonant frequencies that are dependent on the choice of initial guess pulses. GRAPE typically performs a global (all time-slices at once) gradient update of piecewise-constant control amplitudes, while Krotov updates the control field sequentially in time via a local update rule. As a result, while GRAPE can recruit useful resonant components, it also does not penalize non-resonant frequency content, and is thus less efficient in removing the latter than Krotov.
%~\cite{Jager2014BECcontrol}
Despite these differences, GRAPE method is still able to achieve fidelities close to unity for different initial guess pulses, as shown in Fig.~\ref{fig:closed_GR_init_fid}.

% For different choices of initial guess pulses, such as Random~\ref{fig:closed_GR_init}(a), Square~\ref{fig:closed_GR_init}(b), and Zero~\ref{fig:closed_GR_init}(c) pulses, the spectra of optimized pulse show different properties. The peaks in the spectrum are dependent to the initial guess pulses because in this method, the pulse updates during the optimization are locally in time. However, the optimized fidelity is almost 1 in all the cases~\ref{fig:closed_GR_init_fid}(a). The difference of the fidelities when the Zero pulse is taken as the reference is shown in~\ref{fig:closed_GR_init_fid}(b). 

\begin{figure}[ht]
    \centering
    \includegraphics[width=0.6\linewidth]{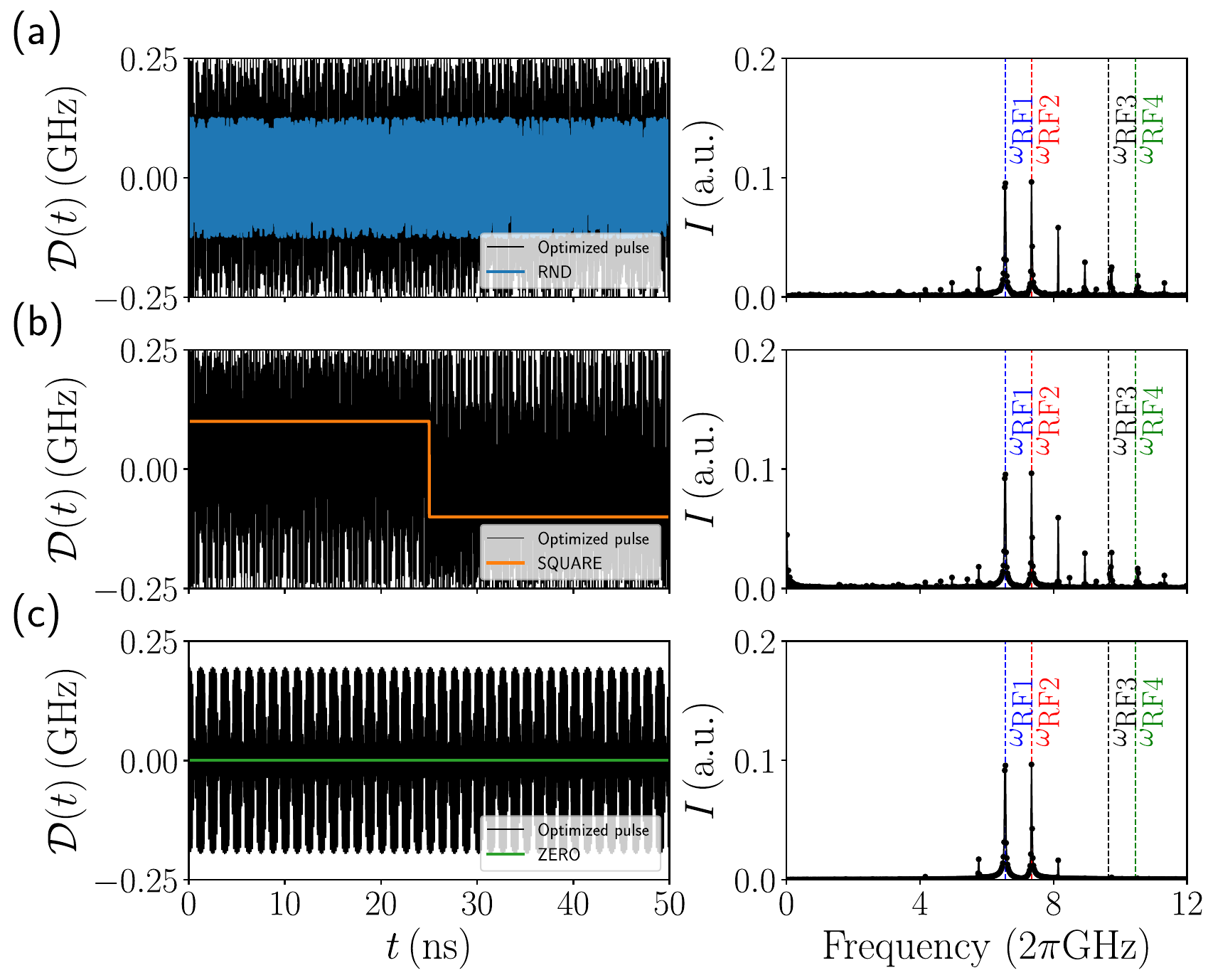}
    \caption{\textbf{Comparison of optimized pulses with between different initial guess pulses.} Three different pulses (a) Random pulse (RND), (b) square pulse (SQUARE), and (c) zero pulse (ZERO) used as initial guess pulses in GRAPE method. All the pulses are optimized for NOT gate on the second qubit. The drift Hamiltonian parameters in the closed system are $\omega_1 = 20\pi~\text{GHz}$, $\omega_2 = 14\pi~\text{GHz}$, and $J = 5~\text{GHz}$. The width is set to $\tau = 50$ ns.}
    \label{fig:closed_GR_init}
\end{figure}

\begin{figure}[ht]
    \centering
    \includegraphics[width=0.6\linewidth]{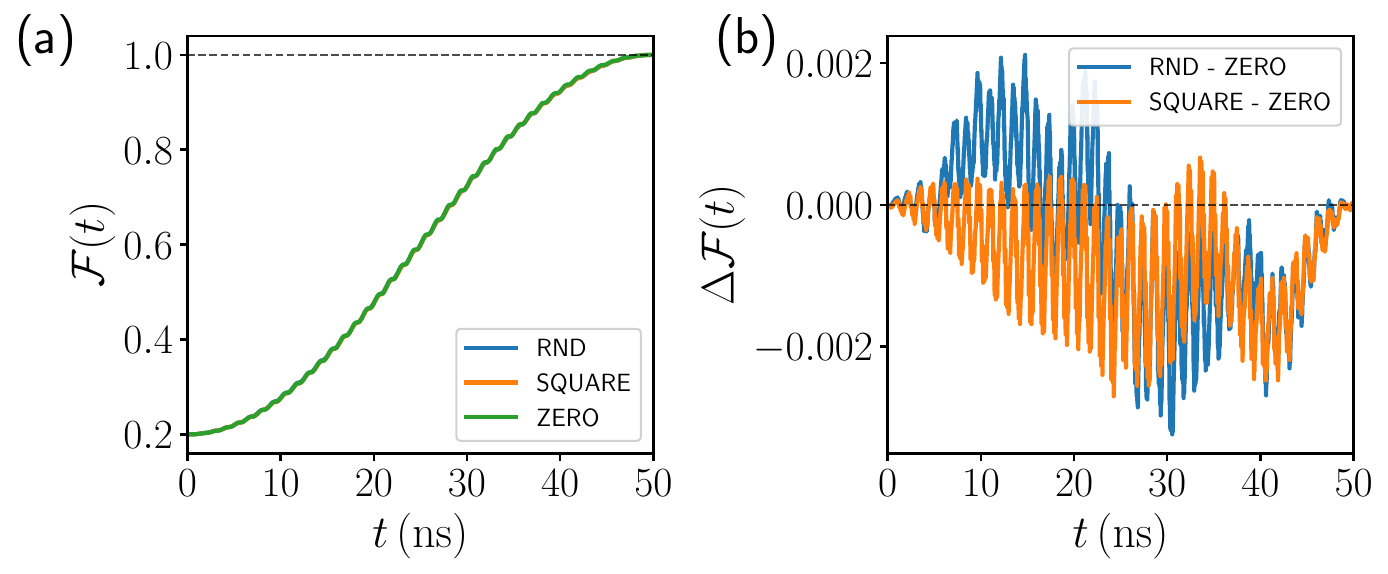}
    \caption{\textbf{Comparison of gate fidelities for different initial guess pulses.}
    (a) Time evolution of the NOT gate fidelity with optimized pulses presented in Fig.~\ref{fig:closed_GR_init}. They all approach unity at the end of the control sequence.
    (b) Fidelity differences with respect to ZERO initial guess pulse, shown as (RND - ZERO) and (SQUARE - ZERO).}
    \label{fig:closed_GR_init_fid}
\end{figure}
}

\clearpage
\section{Krotov results dependency to time resolution}

In this section, we discuss the choice of the time step in the Krotov method simulations. In Fig.~\ref{fig:qoct_no_decoherence}, a fine time step of $0.01$ ns was used for a system with Larmor frequencies $\omega_1 = 20\pi~\text{GHz}$ and $\omega_2 = 14\pi~\text{GHz}$ to ensure an accurate time evolution. Repeating the calculations for the same system with a time step twice as large, $0.02$ ns, yields the same fidelity of 1 (Fig.~\ref{fig:closed_Krotov_tstep02}). These Larmor frequencies were chosen to be well separated, allowing for a clear analysis of the Fourier-transformed spectrum.

% From the theoretical perspective, we choose the time step to be at least 2-time smaller than the intrinsic time scale of the system (e.g. Larmor frequency of 10 GHz ~ intrinsic time of 0.1 ns) to ensure the accurate time evolution of the system. We repeated the calculations with larger time segment 0.05 ns for a closed system of Larmor frequencies taken from the experimental data as we mentioned in section “Limit with realistic parameters” (for two spins of Larmor frequencies 15 GHz and 13 GHz with the J coupling 0.1 GHz). The fidelity is 1 and this result is presented in Fig. S9. The choice of time steps highly depends on the Larmor frequencies. 

% In the previous calculation with the time step 0.01 ns, our choice of Larmor frequencies are large to have well separated frequencies and in order to study the FT spectrum, we need to use a fine time segment. In addition, we could achieve the same fidelity 1 with a two times larger time segment of 0.02 ns which is presented in Fig. S8.

\begin{figure}[ht]
    \centering
    \includegraphics[width=0.6\linewidth]{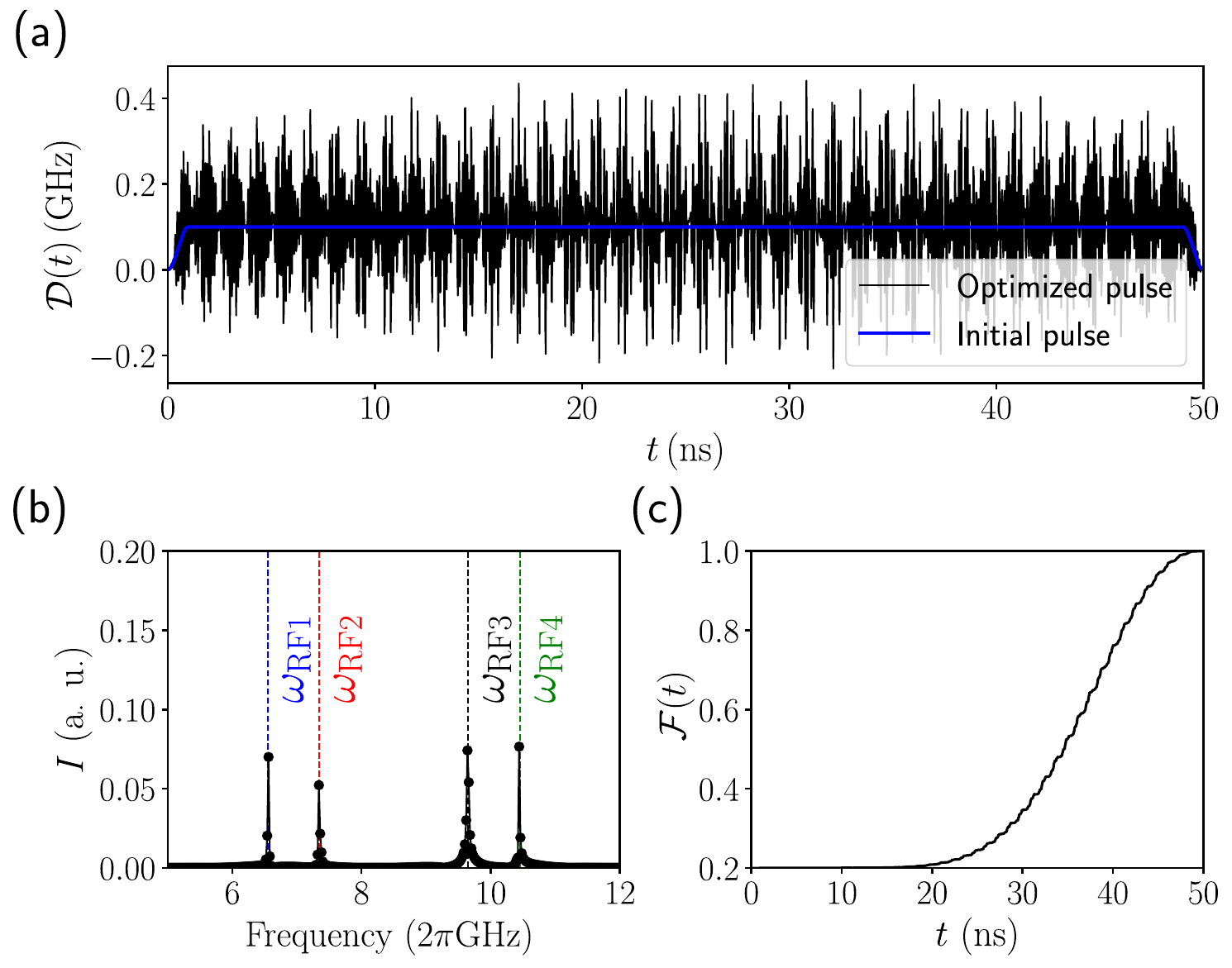}
    \caption{\textbf{Realization of a NOT gate on the second qubit using Krotov method.} (a) The optimal pulse (black) compared to the initial flattop pulse (blue).. The pulse width is set to $\tau = 50$ ns with the time segment $t_{\mathrm{step}}=0.02$ ns. The drift Hamiltonian parameters are $\omega_1 = 20\pi~\text{GHz}$, $\omega_2 = 14\pi~\text{GHz}$, and $J = 5~\text{GHz}$. (b) Fourier transform of the optimal pulse. (c) Fidelity of the NOT gate as a function of time, reaching nearly 1 at the end of the optimal pulse.}

    \label{fig:closed_Krotov_tstep02}
\end{figure}

For a closed system with Larmor frequencies taken from experimental data, $\omega_1 = 15~\text{GHz}$ and $\omega_2 = 13~\text{GHz}$, and a coupling $J = 0.1~\text{GHz}$, as discussed in the section “Limit with realistic parameters” in the main text, the fidelity remains 1 with a time step of $0.05$ ns (Fig.~\ref{fig:closed_Krotov_tstep05}). Although the Larmor frequencies are relatively close, the optimized pulse is still able to find the target evolution and achieve a high fidelity.

\begin{figure}[ht]
    \centering
    \includegraphics[width=0.6\linewidth]{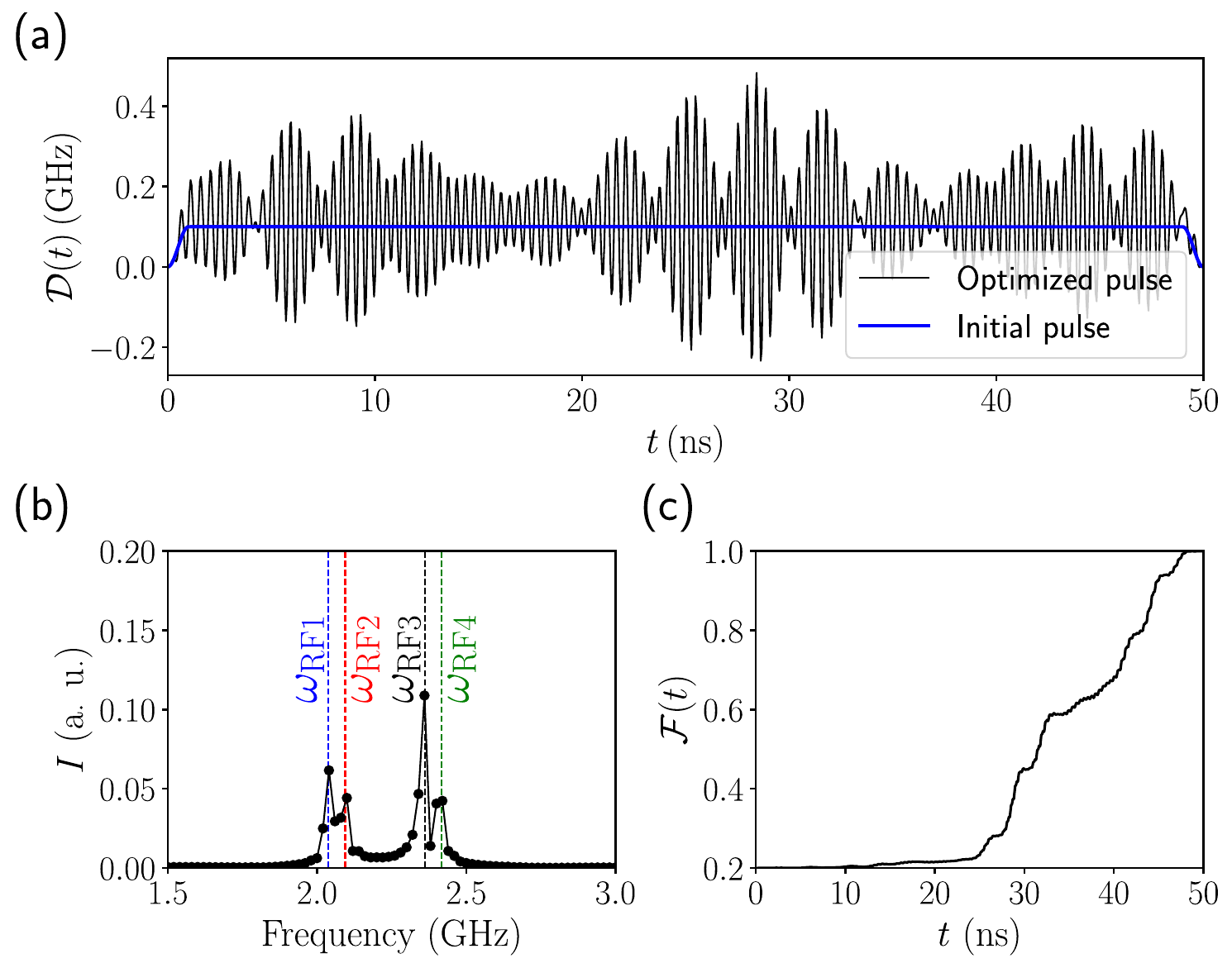}
    \caption{\textbf{Realization of a NOT gate on the second qubit using Krotov  method.} (a) The optimal pulse (black) compared to the initial flattop pulse (blue). The pulse width is set to $\tau = 50$ ns with the time segment $t_{\mathrm{step}}=0.05$ . The drift Hamiltonian parameters are $\omega_1 = 15~\text{GHz}$, $\omega_2 = 13~\text{GHz}$, and $J = 0.35~\text{GHz}$. (b) Fourier transform of the optimal pulse. (c) Fidelity of the NOT gate as a function of time, reaching nearly 1 at the end of the optimal pulse.}

    \label{fig:closed_Krotov_tstep05}
\end{figure}

}

\clearpage
%\section{Effect of time over-/undersampling on the pulses}
{\color{black}
\section{Stability of pulses with respect to the sampling rate and coupling strengths}

\subsection{Impact of the sampling rate}

For the QOCT optimization we chose a time-step smaller than the highest frequency of the system to ensure an accurate time-evolution of the Hamiltonian. Real pulses used in experiments may not achieve such a high time resolution, in this context described as \textit{sampling rate} which is the inverse of the stepsize, $f_s=1/\Delta t$. In the following we discuss how strongly the gate performance depends on the temporal resolution used to represent the optimized control waveform. We define an sampling ratio, that is the time step of the sample relative to the reference time step, i.e. $r=\Delta t/\Delta t^\mathrm{ref}$ with $\Delta t^\text{ref} = 0.01$~ns. We found that over a wide range of oversampling ($r<1$) factors, the average fidelity remains essentially unchanged.
For the reference parameter set, $\mathcal{F}_{\mathrm{avg}}$ stays close to its baseline value (near $0.9$) $r$ up to unity, and then degrades sharply over a comparatively narrow interval. Conversely, the infidelity $1-\mathcal{F}_{\mathrm{avg}}$ rises rapidly by orders of magnitude, see Fig.~\ref{fig:sm_figure15} (a,b).
The vertical dashed lines provide a physically transparent interpretation of this threshold: they indicate $r$-values corresponding to one tenth of the characteristic oscillation periods associated with the dominant transition frequencies $\omega_{\mathrm{RF}1}$--$\omega_{\mathrm{RF}4}$.
The close alignment between the onset of fidelity loss and these markers indicates that the breakdown occurs once the effective sampling becomes too coarse to resolve the fastest relevant dynamics, causing the piecewise-constant representation of the control to distort the intended rotations. The same trend is observed for the simulations related to experiments, see Fig.~\ref{fig:sm_figure15} (c,d). Here, the threshold is shifted, consistent with the modified spectral structure of the system and the shorter pulse duration in this regime. Notably this drop in fidelity appears only for a time grid coarser by almost an order of magnitude than the original timestep.

Taken together, Fig.~\ref{fig:sm_figure15} shows that high-fidelity operation is compatible with substantial over- and undersampling of the optimized pulse, while also providing a practical design rule: the sampling rate must remain comfortably above the largest relevant transition frequency (here summarized by $\omega_{\mathrm{RF}1\text{--}4}$) to preserve gate fidelity.

\begin{figure}[ht]
    \centering
    \includegraphics[width=0.478\textwidth]{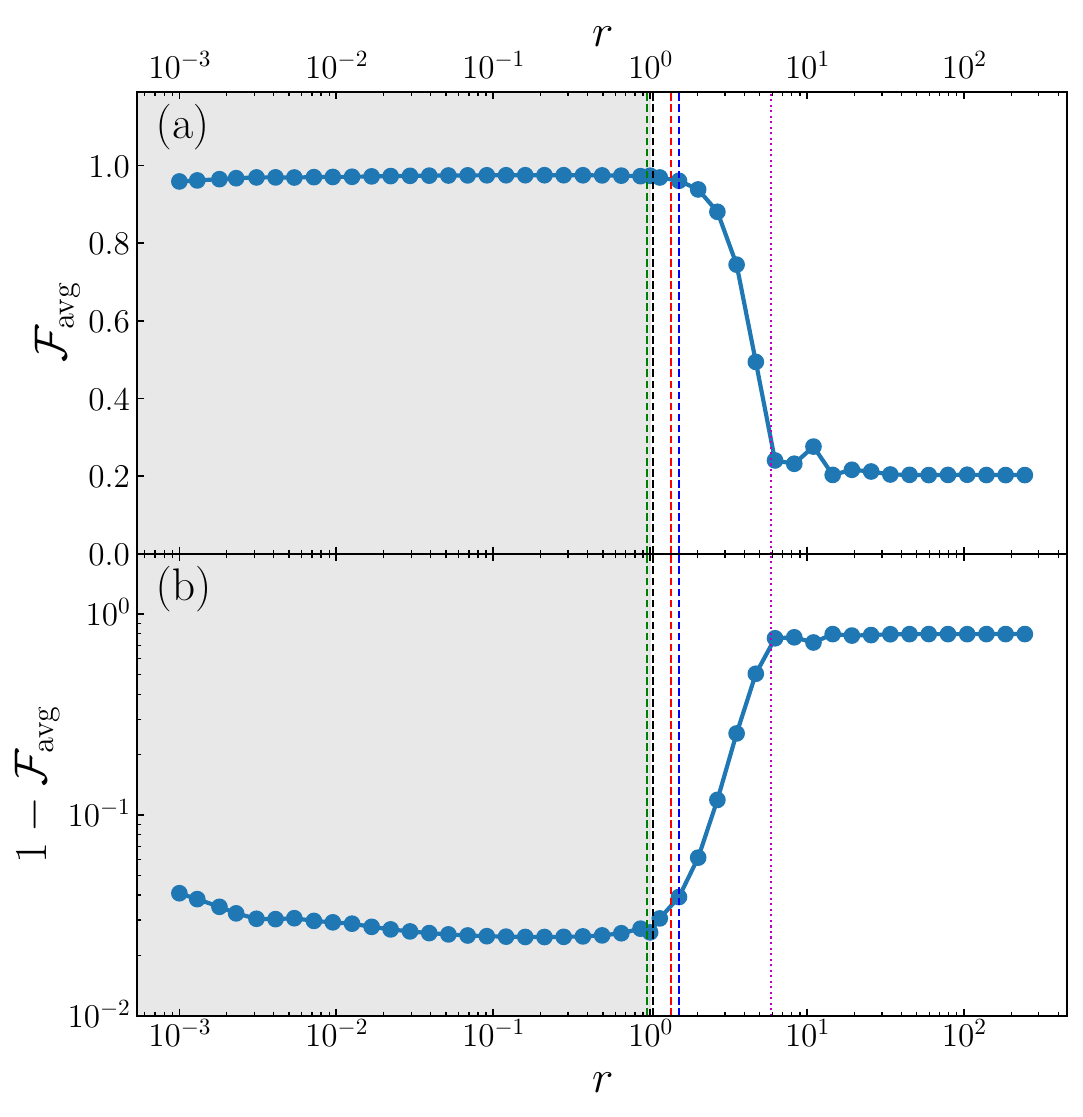}\hfill
    \includegraphics[width=0.502\textwidth]{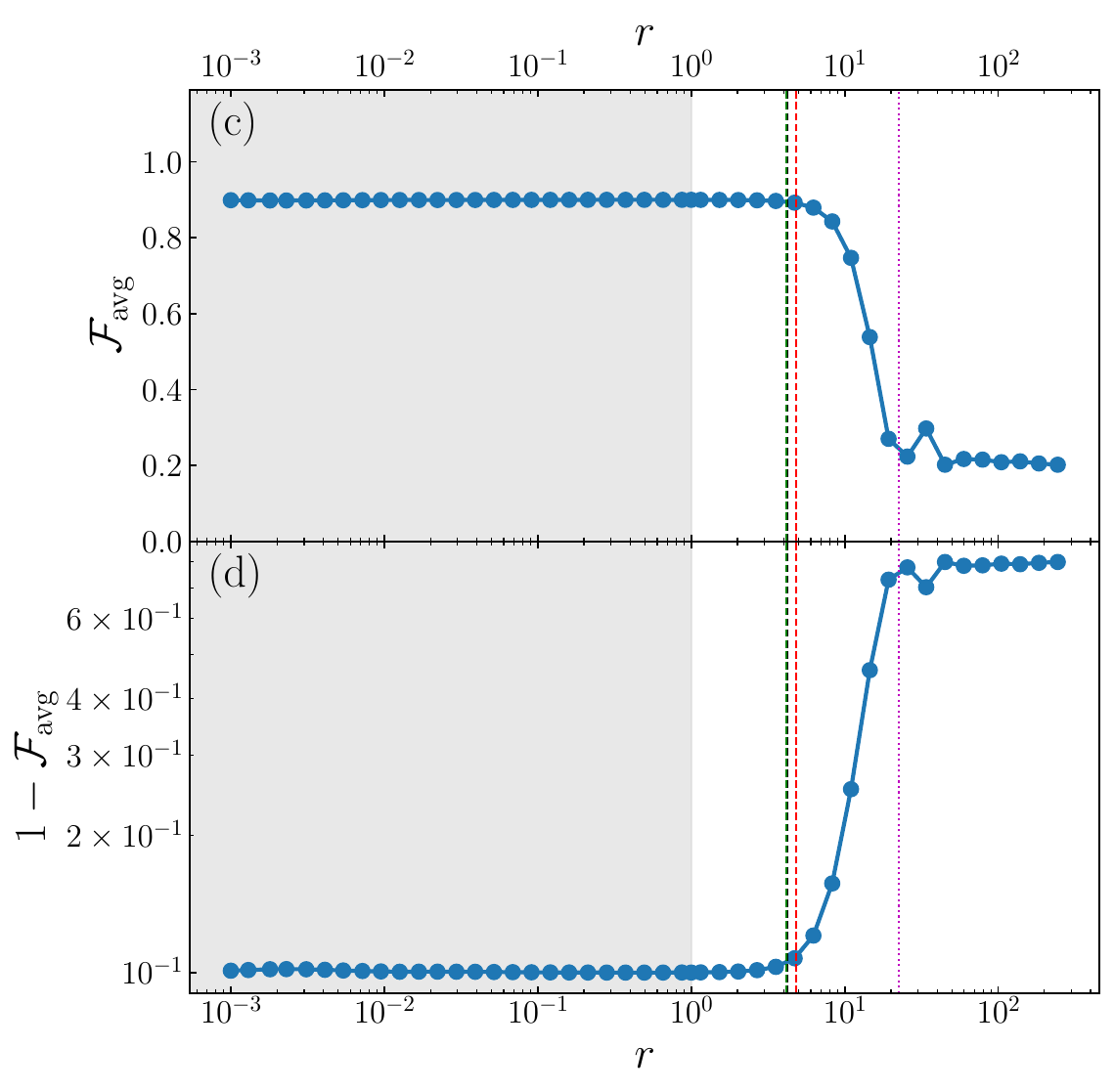}
    \caption{
    (a) Dependence of the average gate fidelity $\mathcal{F}_\mathrm{avg}$ 
    and (b) the corresponding infidelity $1-\mathcal{F}_\mathrm{avg}$ on the pulse undersampling factor $r$
    for a control pulse originally optimized using a time step of $\Delta t^\text{ref} = 0.01\,\mathrm{ns}$
    under Markovian relaxation noise with $T_1^{(1)} = T_1^{(2)} = 1000\,\mathrm{ns}$.
    The undersampling factor on the horizontal axis corresponds to the ratio between the effective
    time step used for pulse discretization and the original optimization time step, such that a
    value of $1$ corresponds to the original resolution and larger values indicate coarser sampling.
    The qubit parameters are $\omega_1 = 20\pi\,\mathrm{GHz}$, $\omega_2 = 14\pi\,\mathrm{GHz}$,
    $J = 5\,\mathrm{GHz}$, and  $\tau=30\,\mathrm{ns}$, which are the main parameters
    used throughout the main text.
    (c,d) Same analysis for a pulse designed using parameters extracted from \cite{Wang2023universal} with $\omega_1 = 15\,\mathrm{GHz}$,
    $\omega_2 = 13\,\mathrm{GHz}$, $J = 0.1\,\mathrm{GHz}$, and 
    $\tau=12.5\,\mathrm{ns}$.
    In all panels, the vertical dashed lines indicate characteristic undersampling values associated
    with one tenth of the four dominant oscillation periods of the system,
    corresponding to the transition frequencies $\omega_{\mathrm{RF}1}$,
    $\omega_{\mathrm{RF}2}$, $\omega_{\mathrm{RF}3}$, and $\omega_{\mathrm{RF}4}$, while the dotted line indicates the spin "flip-flop" rate  $\omega_{1}+\omega_{2}$, finally, the shaded area indicates values corresponding to oversampling ($r<1$).}
    %These markers highlight the onset of fidelity degradation as the pulse sampling rate approaches the characteristic dynamical time scales of the driven system.}
    \label{fig:sm_figure15}
\end{figure}

\subsection{Impact of deviations in the exchange coupling}
We also estimate the stability of optimized fidelity with respect to changes in coupling strength $J$. We first optimized a NOT gate with drift Hamiltonian parametrized by exchange coupling constant $J_0$. Then, we applied the optimized pulse to a system with slight deviations in coupling strength $\Delta J=J-J_0$ but otherwise identical Larmor frequencies. The results shown in Fig.~\ref{fig:stability_J} indicate that the optimized pulses give a reasonably stable fidelity within a window of $\pm 1\%$ change in $J$ for both closed and open systems.

\begin{figure}[ht]
    \centering
    \includegraphics[width=0.5\linewidth]{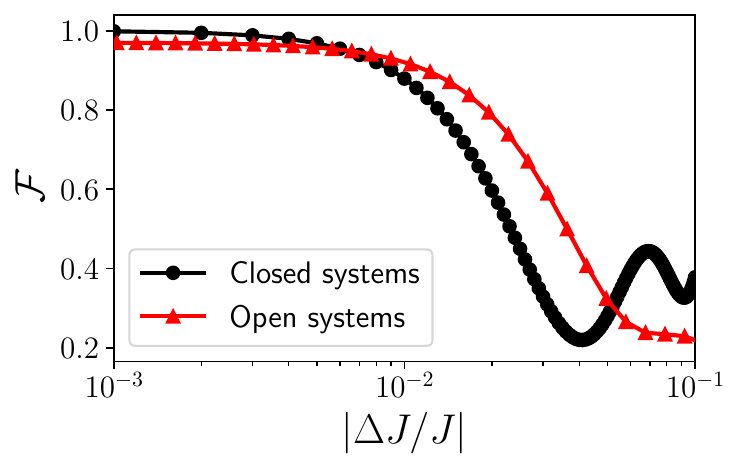}
    \caption{Stability of optimized fidelity with respect to changes in coupling strength $J$ for (a) closed systems and (b) open systems. $\mathcal{F}$ is the optimized gate fidelity of a NOT gate on remote qubit. The parameters of drift Hamiltonian are $\omega_1 = 20\pi\,\mathrm{GHz}$, $\omega_2 = 14\pi\,\mathrm{GHz}$,
    $J = 5\,\mathrm{GHz}$, and the pulse width is $\tau=30\,\mathrm{ns}$. For the open systems, $T_1^{(1)} = T_1^{(2)} = 1000~\text{ns}$.}
    \label{fig:stability_J}
\end{figure}

}

\clearpage
\section{Collapse operators for Lindblad master equation}

In the real physical systems, decoherence processes induced from the coupling of spins with the substrate and tunneling current imposes significant limitations on the spin qubits. We use the Linblad formalism to account for the decoherence and the master equation takes the following form 
\begin{equation}
\frac{d\rho}{dt} = -\frac{i}{\hbar} \left[ H(t), \rho \right] + \sum_k \left( \mathcal{L}_k \rho \mathcal{L}_k^\dagger - \frac{1}{2} \mathcal{L}_k^\dagger \mathcal{L}_k \rho - \frac{1}{2} \rho \mathcal{L}_k^\dagger \mathcal{L}_k \right),
\end{equation}
where $\rho$ is the density matrix of the system, and $\mathcal{L}_k = \mathcal{L}^\text{relax}_k + \mathcal{L}^\text{dephase}_k$ are $k$-th channel collapse operators to model energy relaxation and pure dephasing.

As for the pure dephasing process, the standard operators are written via the pure dephasing time as follows
\begin{equation}
    \mathcal{L}^\text{dephase}_k = \sqrt{\frac{1}{2 T_{\phi}^{(k)}}} \hat{\sigma}_z^k.
\end{equation}
For each spin $k$, we have the following identity
\begin{equation}
    \frac{1}{T_2^{(k)}} = \frac{1}{2 T_1^{(k)}} + \frac{1}{T_\phi^{(k)}}.
\end{equation}
In our previous study~\cite{Phark2023DoubleResonance}, all experiments were reproduced using the energy relaxation only, thus we employ the theoretical upper limit of the coherence time, i.e. $T_2 = 2 T_1$.
%(cite Delgado, Ternes papers here)~\cite{Ternes2015, Delgado2017}

\clearpage
\section{Spectrum of optimized pulses for different update parameters of the Krotov method}

\begin{figure}[h!]
    \centering
    \includegraphics[width=0.5\linewidth]{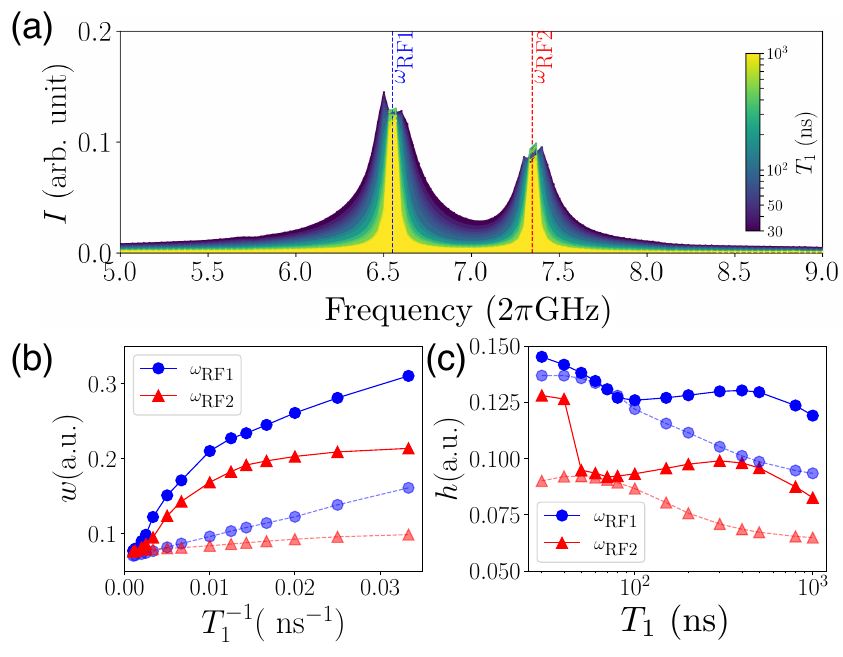}
    \caption{(a) Spectrum of optimized pulses with $\lambda_a = 0.1$. (b)(c) Width and height of the peaks with $\lambda_a = 0.5$ are shown with less opaque markers and dashed lines for comparison. Other parameters are identical to those in Fig.~\ref{fig:spectrum_vs_relaxation} in the main text: pulse width $\tau = 30$ ns, drift Hamiltonian is with $\omega_1 = 20 \pi \text{ GHz}, \omega_2 = 14 \pi \text{ GHz}, J = 5 \text{ GHz}$. }
    \label{fig:spectrumdifferentlambda}
\end{figure}

In order to gauge the importance of the update parameter $\lambda_a$ in the Krotov  method, we performed optimizations with different values of $\lambda_a$. The results are shown in Fig.~\ref{fig:spectrumdifferentlambda}. We observe that the evolution of the width and height of the peaks as a function of $T_1$ is highly dependent on the choice of $\lambda_a$. For smaller values of $\lambda_a$, the peaks width increases more rapidly as a function of $1/T_1$, when compared to the case with larger $\lambda_a$ shown in the main text, until it eventually saturates. Moreover, the peak heights behave non-monotonically as a function of $T_1$ for smaller $\lambda_a$, while they decrease monotonically for the larger $\lambda_a$ chosen  in the main text. This indicates that the choice of $\lambda_a$ can significantly influence the characteristics of the optimized pulses, particularly in the presence of relaxation effects, and should be carefully considered when designing control protocols for quantum systems using the Krotov method, the desired bandwidth being one of the main factors to take into account.

\clearpage
\section{Spectrum of optimized pulses with the Rabi rate synchronized pulse as an initial guess}

\begin{figure}[ht]
    \centering
    \includegraphics[width=0.5\linewidth]{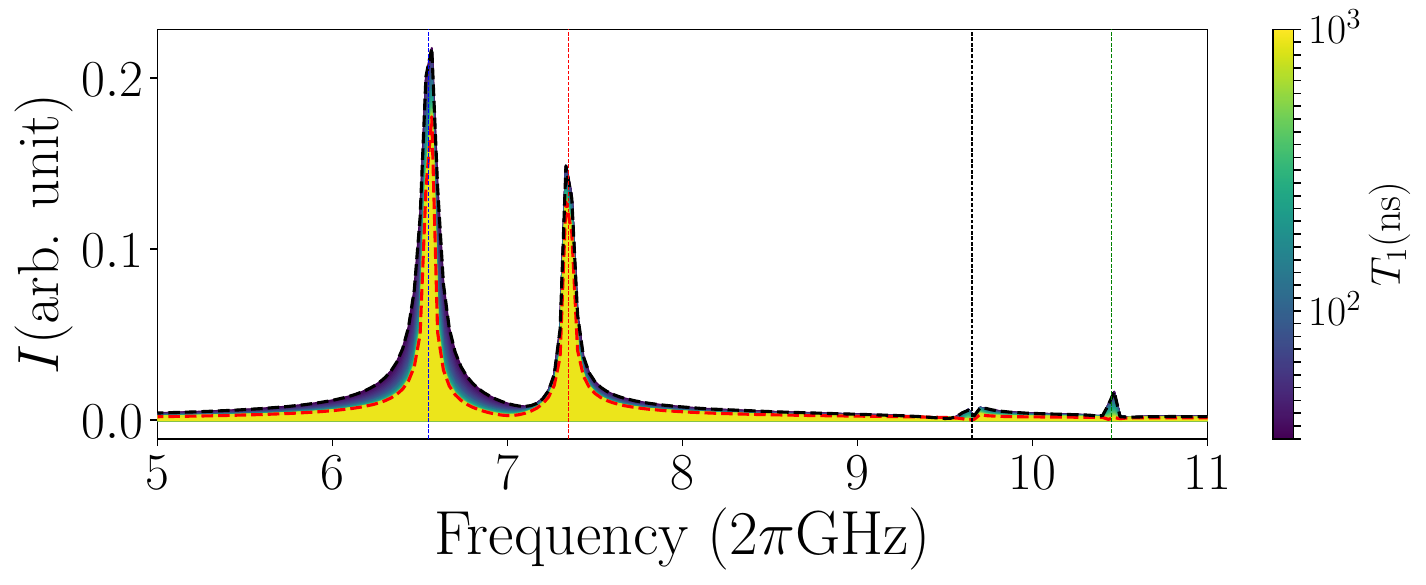}
    \caption{Spectrum of the resulting pulses
    after 100 iterations of Krotov  method
    optimizing for open system dynamics, for
    increasing $T_1^{(1)} = T_1^{(2)}$. The
    initial guess, whose spectrum is plotted in
    red, was taken to be the result of a
    closed-system Krotov optimization for 500
    iterations, the latter's guess pulse being the
    Rabi synchronized pulse. The parameters of
    drift Hamiltonian are $\omega_1 = 20 \pi
    \text{ GHz}, \omega_2 = 14 \pi \text{ GHz}$,
    $J = 5 \text{ GHz}$. The pulse width is $\tau
    = 30$ ns.}
    \label{fig:spectrum_synchronized}
\end{figure}

\begin{figure}[ht]
    \centering
    \includegraphics[width=0.5\linewidth]{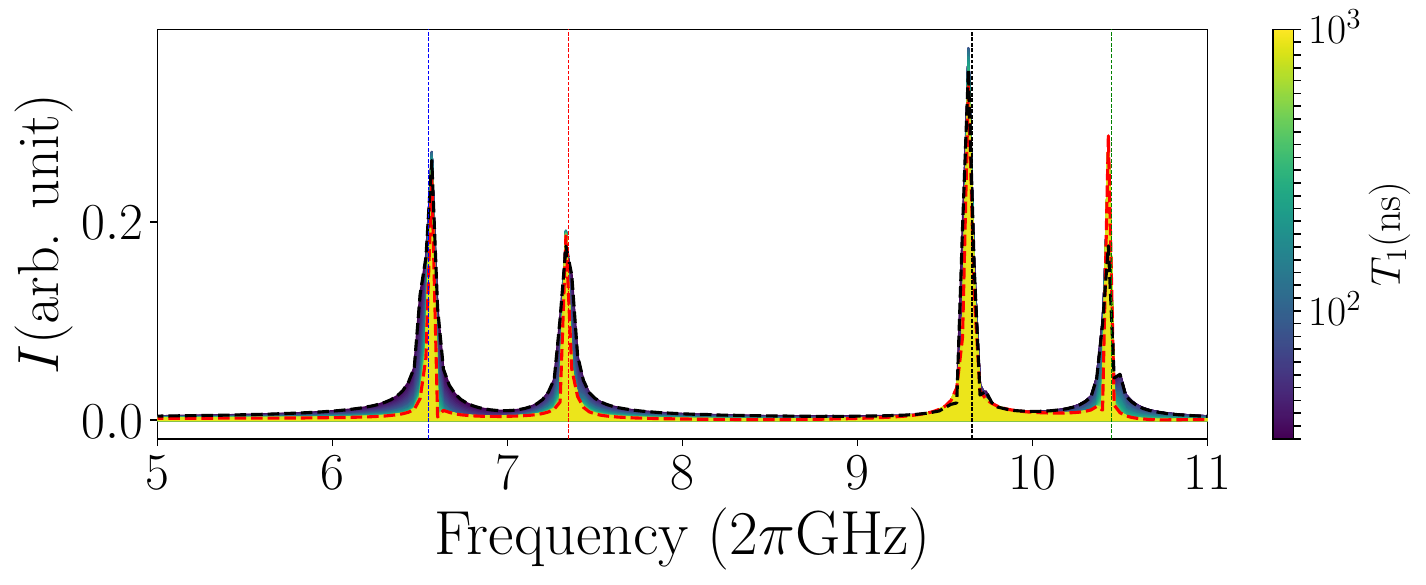}
    \caption{Spectrum of the resulting pulses
    after 100 iterations of Krotov  method
    optimizing for open system dynamics, for
    increasing $T_1^{(1)} = T_1^{(2)}$. The
    initial guess, whose spectrum is plotted in
    red, was taken to be the result of a
    closed-system Krotov optimization for 500
    iterations, the latter's guess pulse being the
    flattop-seeded optimized pulse in the closed system. The parameters of
    drift Hamiltonian are $\omega_1 = 20 \pi
    \text{ GHz}, \omega_2 = 14 \pi \text{ GHz}$,
    $J = 5 \text{ GHz}$. The pulse width is $\tau
    = 30$ ns.}
    \label{fig:spectrum_flattop}
\end{figure}

To further investigate the influence of the initial guess on the optimized pulses, we employed the Rabi rate synchronized pulse as an initial guess for the Krotov optimization targeting open system dynamics. Compared to the flattop pulse used in the main text, the Rabi rate synchronized pulse has a narrower spectrum, as it is a simple sinusoidal pulse with two frequency components already on resonance with the qubits. The results are shown in Fig.~\ref{fig:spectrum_synchronized}. We observe that the optimized pulses retain a spectrum similar to that of the initial guess, with only two dominant peaks corresponding to the qubit frequencies. This contrasts with the results obtained using the flattop pulse as an initial guess shown in Fig.~\ref{fig:spectrum_flattop}, where four peaks were observed. This suggests that the choice of initial guess can significantly influence the characteristics of the optimized pulses, particularly in terms of their spectral content. The Rabi rate synchronized pulse, being closer to the desired resonant frequencies, leads to optimized pulses that maintain this simplicity in their spectral profile, however, in the closed system case, the optimization starting from the Rabi synchronized pulse does not reach unit fidelity, as the initial guess is not sufficiently close to the optimized solution and close to a local minimum. This is why we used instead the result of a closed-system optimization starting from the flattop pulse as an initial guess for the open-system optimization, which, in the closed system case, reaches unit fidelity.

\clearpage
\section{Performance under noise of noise-informed optimization vs. noise-agnostic (closed system) optimization}

\begin{figure}[ht]
    \centering
    \includegraphics[width=.5\linewidth]{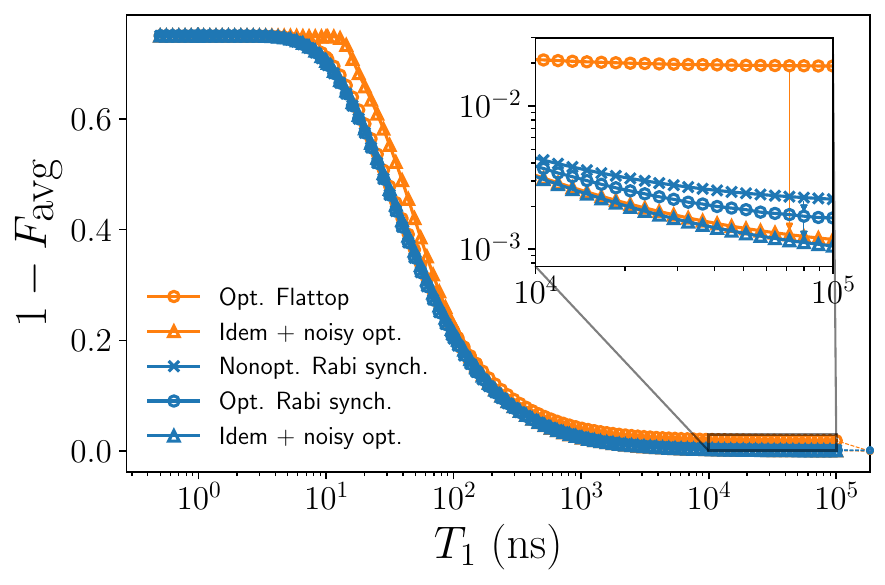}
    \caption{Infidelities of different pulses
    resulting from Krotov optimizations, for
    increasing $T_1^{(1)} = T_1^{(2)}$. The values
    for the pulse resulting from 500 iterations of
    a closed-system Krotov optimization with the
    Rabi synchronized (resp. flattop) pulse as an
    initial guess is shown in blue (resp. orange) with round markers,
    while the values for the pulses optimized for
    open-system dynamics with the latter
    closed-system-optimized pulses as initial
    guesses are shown in blue (resp. orange) with triangular markers. Moreover, the results of the naive Rabi synchtronized pulse with no optimization whatsoever is shown with cross markers. The
    parameters of drift Hamiltonian are $\omega_1
    = 20 \pi \text{ GHz}, \omega_2 = 14 \pi \text{
    GHz}$, $J = 5 \text{ GHz}$. The pulse width is
    $\tau = 30$ ns.}
    \label{fig:optimal_works}
\end{figure}

Fig.~\ref{fig:optimal_works} compares the fidelity of pulses obtained by (i) optimizing in a closed system and then using the resulting pulse in a dissipative setting (``noise-agnostic''), and (ii) re-optimizing the same pulse directly on the open-system dynamics (``noise-informed''). The scan is performed versus a common relaxation time, with $T_1^{(1)}=T_1^{(2)}$, while keeping the drift and control parameters fixed. For each of two natural seeds, a flattop pulse and a Rabi-rate–synchronized pulse, we first run a closed-system Krotov optimization and then re-optimize under noise starting from that closed-system solution.

Across the full $T_1$ range, noise-informed optimization consistently reduces the infidelity relative to deploying a closed-system–optimized pulse in a dissipative device. As $T_1\!\to\!\infty$ the curves coalesce, while for very short $T_1$ all pulses saturate at a finite error set by the limited lifetime during the fixed gate time $\tau$ (and by the chosen gate). In the long-$T_1$ regime, the noise-agnostic pulse initialized from a flattop seed performs markedly worse than the others, for example it performs even worse than the nonoptimized naive Rabi synchronized pulse; re-optimizing under noise, however, yields a substantial fidelity gain that brings it in line with the best performer in this regime: the noise-informed pulse initialized from the Rabi-synchronized seed. The noise-agnostic counterpart of the latter still trails its noise-informed version, but it clearly outperforms the flattop-seeded, noise-agnostic pulse. By contrast, in the intermediate-noise regime ($T_1\!\sim\!10 - 100\,\mathrm{ns}$), the noise-informed pulse initialized from a flattop seed becomes the worst of the set, even underperforming its own noise-agnostic counterpart. This indicating a failure mode of the optimization. Notably however, this occurs where absolute fidelities are already low ($\sim 0.5$), a range that is not suitable for coherent quantum control.

This trend is consistent with our modeling and control framework: the open-system dynamics are described by a GKSL master equation with relaxation (and, when used, dephasing) included through collapse operators, so that the optimization can actively reshape spectral weight and timing to mitigate irreversible loss rather than merely cancel coherent errors. The practical takeaway is that Krotov pulses should be \emph{optimized on the open-system model} to encode robustness to relaxation, choosing the seed to respect experimental bandwidth while relying on the noise-informed step to recover fidelity.

% \bibliography{ref.bib}
% \end{document}

\end{document}